\documentclass[12pt]{article}
\usepackage[letterpaper,top=2cm,bottom=2cm,left=2cm,right=2cm,marginparwidth=1.75cm]{geometry}
\pdfoutput=1

\usepackage{graphicx} 
\usepackage{amsmath}
\usepackage{amsfonts}
\usepackage{amssymb}
\usepackage{amsthm}
\usepackage{enumerate}
\usepackage{appendix}
\usepackage{authblk}
\usepackage[font=small,labelfont=bf]{caption}
\usepackage[font=small,labelfont=bf]{subcaption}
\usepackage{tikz}
\usetikzlibrary{shapes.geometric}
\usepackage{tabularx}
\usepackage{ifthen}
\usepackage{physics}
\usepackage{dsfont}
\usepackage{mathdots}
\usepackage{hyperref}
\usepackage{todonotes}
\usepackage[normalem]{ulem}

\newcommand{\hgproduct}{\boxtimes}

\renewcommand{\op}[1]{\operatorname{#1}}

\definecolor{qubitcolor}{rgb}{1,1,1}
\definecolor{xcheckcolor}{rgb}{0.89,0,0.13}
\definecolor{zcheckcolor}{rgb}{0,0.5,1}
\definecolor{classicalbitcolor}{rgb}{1,1,1}
\definecolor{classicalcheckcolor}{rgb}{0,0,0}

\newtheorem{lemma}{Lemma}
\newtheorem{example}{Example}
\newtheorem{proposition}{Proposition}

\newtheorem{definition}{Definition}
\newtheorem{remark}{Remark}
\newtheorem{theorem}{Theorem}

\title{Pruning qLDPC codes: Towards bivariate bicycle codes with open boundary conditions}
\author[1]{Jens Niklas Eberhardt\thanks{mail{@}jenseberhardt.com}}
\author[2]{Francisco Revson F. Pereira\thanks{francisco.revson{@}meetiqm.com}}
\author[2]{Vincent Steffan\thanks{vincent.steffan{@}meetiqm.com}}
\affil[1]{Mathematical Institute of the University of Bonn, Germany}
\affil[2]{IQM Quantum Computers, Germany}

\begin{document}

\maketitle

\begin{abstract}
Quantum low-density parity-check codes are promising candidates for quantum error correcting codes as they might offer more resource-efficient alternatives to surface code architectures. In particular, bivariate bicycle codes have recently gained attention due to their 2D-local structure, high encoding rate, and promising performance under simulation. In this work, we will explore how one can transform bivariate bicycle codes defined on lattices with periodic boundary conditions to codes with the same locality properties on a 2D lattice with open boundary conditions. For this, we introduce the concept of pruning quantum codes. We explain how pruning bivariate bicycle codes is always possible when the codes are hypergraph products of two classical cyclic codes. We also indicate that this might be possible for more general bivariate bicycle codes by constructing explicit examples. Finally, we investigate fault-tolerant quantum computation using the constructed pruned codes by describing fold-transversal gates. 
\end{abstract}

\section{Introduction}

Fault-tolerant quantum computation will require the usage of quantum error-correcting codes to suppress errors~\cite{Knill_1998,aharonov1999faulttolerant}. One promising candidate is the surface code~\cite{bravyi1998quantum} which, with its 2D-planar nearest neighbor connectivity, is particularly suitable for implementation on near-term devices, for example superconducting qubit architectures. A downside of the surface code is the large overhead per logical qubit.

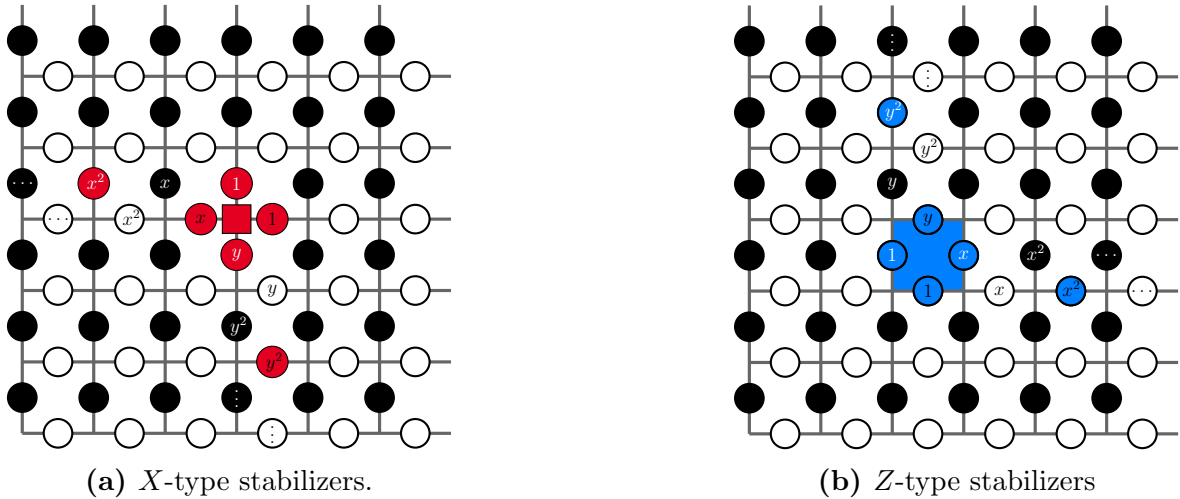
\begin{figure}
     \centering
     \begin{subfigure}[b]{0.45\textwidth}
         \centering
          
 \begin{tikzpicture}[scale = .95]
        \foreach \x in {1,2,3,4,5,6} (
            \draw[very thick, color = black!60!white] (\x,0) -- (\x, 6);
        )
        \foreach \x in {0,1,2,3,4,5} (
            \draw[very thick, color = black!60!white] (1,\x) -- (7,\x);
        )
        \draw[fill = xcheckcolor] (3.8,2.8) rectangle ++(.4,.4);
        \foreach \x in {1,2,3,4,5,6} (
            \foreach \y in {0,1,2,3,4,5} (
                \draw[thick, fill = white] (\x + .5, \y) circle (0.2);
            )
        )
        \foreach \x in {1,2,3,4,5,6} (
            \foreach \y in {1,2,3,4,5,6} (
                \draw[thick, fill = black] (\x , \y - .5) circle (0.2);
            )
        )
        \draw[fill = xcheckcolor] (4,3.5) circle (0.22);
        \draw[fill = xcheckcolor] (4,2.5) circle (0.22);
        \draw[fill = xcheckcolor] (2,3.5) circle (0.22);

        \draw[fill = xcheckcolor] (3.5,3) circle (0.22);
        \draw[fill = xcheckcolor] (4.5,3) circle (0.22);
        \draw[fill = xcheckcolor] (4.5,1) circle (0.22);
        
    \pgfmathsetmacro{\scaler}{.6}
        
        \node[color = white,scale = \scaler] at (4,3.5) {$1$};
        \node[color = white,scale = \scaler] at (3,3.5) {$x$};
        \node[color = white,scale = \scaler] at (2.03,3.525) {$x^2$};
        \node[color = white,scale = \scaler] at (1.03,3.5) {$\dots$};

        \node[color = white,scale = \scaler] at (4,2.5) {$y$};
        \node[color = white,scale = \scaler] at (4.03,1.5) {$y^2$};
        \node[color = white,scale = \scaler] at (4,0.56) {$\vdots$};

        \node[scale = \scaler] at (4.5,3) {$1$};
        \node[scale = \scaler] at (3.5,3) {$x$};
        \node[scale = \scaler] at (2.53,3.025) {$x^2$};
        \node[scale = \scaler] at (1.53,3) {$\dots$};

        \node[scale = \scaler] at (4.5,2) {$y$};
        \node[scale = \scaler] at (4.53,1) {$y^2$};
        \node[scale = \scaler] at (4.5,0.06) {$\vdots$};
        
 \end{tikzpicture}  
         \caption{$X$-type stabilizers.}
         \label{fig:three sin x}
     \end{subfigure}
     \hfill
     \begin{subfigure}[b]{0.45\textwidth}
         \centering
          \begin{tikzpicture}[scale = .95]
\draw[fill = zcheckcolor] (3,2) rectangle ++(1,1);

        \foreach \x in {1,2,3,4,5,6} (
            \draw[very thick, color = black!60!white] (\x,0) -- (\x, 6);
        )
        \foreach \x in {0,1,2,3,4,5} (
            \draw[very thick, color = black!60!white] (1,\x) -- (7,\x);
        )
        \foreach \x in {1,2,3,4,5,6 } (
            \foreach \y in {0,1,2,3,4,5} (
                \draw[thick, fill = white] (\x + .5, \y) circle (0.2);
            )
        )
        \foreach \x in {1,2,3,4,5,6} (
            \foreach \y in {1,2,3,4,5,6} (
                \draw[thick, fill = black] (\x , \y - .5) circle (0.2);
            )
        )

    \draw[thick, fill = zcheckcolor] (4,2.5) circle (0.2);
    \draw[thick, fill = zcheckcolor] (3,2.5) circle (0.2);
    \draw[thick, fill = zcheckcolor] (3,4.5) circle (0.2);

    \draw[thick, fill = zcheckcolor] (3.5,2) circle (0.2);
    \draw[thick, fill = zcheckcolor] (5.5,2) circle (0.2);
    \draw[thick, fill = zcheckcolor] (3.5,3) circle (0.2);

    \pgfmathsetmacro{\scaler}{.6}

        \node[color = white,scale = \scaler] at (3,2.5) {$1$};
        \node[color = white,scale = \scaler] at (4,2.5) {$x$};
        \node[color = white,scale = \scaler] at (5.01,2.54) {$x^2$};
        \node[color = white,scale = \scaler] at (6.02,2.5) {$\dots$};
        
        \node[color = white,scale = \scaler] at (3,3.5) {$y$};
        \node[color = white,scale = \scaler] at (3.02,4.52) {$y^2$};
        \node[color = white,scale = \scaler] at (3,5.56) {$\vdots$};

        \node[scale = \scaler] at (3.5,2) {$1$};
        \node[scale = \scaler] at (4.5,2) {$x$};
        \node[scale = \scaler] at (5.51,2.04) {$x^2$};
        \node[scale = \scaler] at (6.52,2) {$\dots$};

        \node[scale = \scaler] at (3.5,3) {$y$};
        \node[scale = \scaler] at (3.52,4) {$y^2$};
        \node[scale = \scaler] at (3.5,5.06) {$\vdots$};
        
 \end{tikzpicture} 
         \caption{$Z$-type stabilizers}
         \label{fig:five over x}
     \end{subfigure}
        \caption{The $X$-type and $Z$-type stabilizers of the bivariate bicyclic code associated with the polynomials $A(x,y) = 1 + x + y^2, B(x,y) = 1 + y + x^2$. The white qubits are the \textit{horizontal} qubits, the black ones the \textit{vertical} qubits.  The colored qubits are the support of the marked stabilizer. The other polynomial terms sketch how the $X$- and $Z$-type stabilizers look in general. \\
        For the $X$-type stabilizer depicted in (a), the terms of $A(x,y)$ indicate the support of the stabilizer on horizontal qubits (where the polynomial terms are in black font), the terms of $B(x,y)$ the vertical support (in white font). For the $Z$-type stabilizer shown in (b) it is the opposite: The terms of $A(x,y)$ indicate the vertical qubits on which the stabilizer is supported, the terms of $B(x,y)$ on which horizontal qubits the stabilizer is supported.
        }\label{fig:bivariatebicyclestabs}
\end{figure}

Because of that, quantum low-density parity-check codes~\cite{breuckmannQuantumLowDensityParityCheck2021a} have recently gained a lot of attention, in particular, the so-called \textit{bivariate bicycle codes} introduced in~\cite{PhysRevA.88.012311}. Bivariate bicycle codes are generalizations of  \textit{hypergraph product codes}~\cite{Tillich_2014} of pairs of (classical) cyclic codes which, themselves, are natural generalizations of the toric code. We review the construction of these codes in more detail in Section~\ref{sec:background} and only give a brief intuition here. Bivariate bicyclic codes are CSS codes arising from a pair of bivariate polynomials $A(x,y)$ and $B(x,y)$ and a pair of integers $\ell$ and $m$. The qubits of a bivariate bicycle code can be imagined to sit on the edges of a 2D lattice where the integers $\ell$ and $m$ specify the horizontal and vertical size of this lattice, respectively. The polynomials $A(x,y)$ and $B(x,y)$ specify $X$-checks and $Z$-checks for the vertices and plaquettes of the lattice, see Figure~\ref{fig:bivariatebicyclestabs} for a visualization. From Figure~\ref{fig:bivariatebicyclestabs} we can see that the maximum of the degrees of $A(x,y)$ and $B(x,y)$ essentially specifies how non-local checks are.

In~\cite{bravyi2023highthreshold}, some moderate-size examples of bivariate bicycle codes are introduced that require around ten times fewer physical qubits per logical qubits while having a similar code distance and performance under simulation. This better encoding rate is made possible by allowing for slightly higher connectivity with also a higher connection length than surface codes.

A major issue of the bivariate bicycle codes is that they are by construction defined on lattices with \emph{periodic} boundary conditions. Attempts to circumvent this problem include using flip-chip technology~\cite{bravyi2023highthreshold} or folding the code two times and thereby trading the periodic boundary conditions with a slightly higher non-locality~\cite{poole2024architecture}.

In this work, we are interested in transforming bivariate bicycle codes to codes on lattices with open boundary conditions without the need to increase the non-locality of the stabilizer generators. More precisely, we ask for what we call \textit{pruned bivariate bicycle codes}. 
For a stabilizer code $\mathcal{Q}$ induced by stabilizer generators $S_1 , \dots , S_{\ell}$, we call any stabilizer code that arises by deleting certain qubits and stabilizers from $\mathcal{Q}$ such that the remaining stabilizer generators still commute a \textit{pruned code derived from $\mathcal{Q}$}.

If a stabilizer generator $S_i$ is not being removed, but a qubit in its support is, the support is being adjusted accordingly. Clearly, pruning a code does not increase the non-locality of the stabilizer checks. We are interested in pruning bivariate bicycle codes whose stabilizer generators have the same locality properties as the ones of their parent but on a 2D lattice with open boundary conditions. Our first result concerns bivariate bicycle codes that are hypergraph product codes, that is, bivariate bicycle codes associated with univariate polynomials. Note that the logical dimension of such a code is always even, see Section~\ref{sec:background} for an explanation.

\begin{theorem}
Let $\mathcal{Q}$ be a bivariate bicycle code associated with univariate polynomials $A(x,y) = A(x)$ and $B(x,y) = B(y)$ of degrees $r_A$ and $r_B$, respectively, and say $\mathcal{Q}$ has parameters $[[2\ell m, 2k, d]]$. If $A(x) | (x^{\ell} -1)$ and $B(y) | (y^m -1)$, then there exists a pruned bivariate bicycle code $\tilde{\mathcal{Q}}$ with parameters $[[\leq  2\ell m, k, d]]$ that is as local as $\mathcal{Q}$ on a lattice with open boundary conditions.
\end{theorem}
In other words, the trade-off for having open boundary conditions is half of the logical qubits, the number of physical qubits stays roughly the same and the distance stays exactly the same. The exact number of physical qubits of the pruned bivariate bicycle code mainly depends on the degree of the polynomials $A(x,y)$ and $B(x,y)$, for details see Section~\ref{sec: towardsopenboundary}.  We mention that similar ideas have been discussed in~\cite{pecorari2024highrate}.

It turns out that bivariate bicycle codes that are not hypergraph product codes, that is, that are associated with bivariate polynomials, can offer more promising code parameters~\cite{PhysRevA.88.012311,bravyi2023highthreshold}. It is therefore an obvious question if these codes can also be pruned. We show that this is the case, for example, for bivariate bicycle codes associated with $A(x,y) = 1 + x + xy$ and $B(x,y) = 1 + y + xy$ for $l$ and $m$ multiples of $3$. These turn out to give the family of (6.6.6) honeycomb color codes~\cite{Bombin_2006}. For this family of codes, one can find pruned versions that are as local with respect to open boundary conditions and similar code parameters. We also show ways of pruning the bivariate bicycle code associated to the polynomials  $A(x,y) = 1 + x + y^{-1} + xy$ and $B(x,y) = 1 + y +xy + x^{-1}y^{-1}$. We leave the general question for future work.

Recently, fault-tolerant implementations of Clifford gates on bivariate bicycle codes using so-called \textit{fold-transversal gates} have been investigated~\cite{bravyi2023highthreshold, eberhardt2024logicaloperatorsfoldtransversalgates}. These are unitaries which are products of single and two-qubit gates acting on the orbits of so-called $ZX$-dualities. In Section~\ref{sec: foldtransversal}, we briefly review these gates and explain how some of them induce valid logical gates for the pruned codes constructed in Section~\ref{sec: towardsopenboundary}.

This work is structured as follows. In Section~\ref{sec:background}, we introduce the necessary notions from the theory of classical cyclic codes and quantum CSS codes, and briefly explain hypergraph product codes and bivariate bicycle codes. In Section~\ref{sec: towardsopenboundary}, we explain how one can find pruned bivariate bicycle codes and show in Section~\ref{sec: foldtransversal} how so-called fold-transversal gates carry over from the bivariate bicycle codes to their pruned versions. Finally, we explain the open ends of this work in Section~\ref{sec: outlook}.

\subsection*{Acknowledgements}
We thank Shin Ho Choe, Martin Leib, Pedro Parrado, and Fedor \v{S}imkovic for helpful discussions.

\section{Background}\label{sec:background}
We recall some basic concepts from classical and quantum error correction. Though most of what follows can be generalized to non-binary classical codes and qudit quantum codes, we will focus on the binary and qubit case, respectively. 
\subsection{Classical cyclic codes}
A \textit{(classical) linear code} encoding $k$ bits into $n$ bits is a $k$-dimensional subspace $\mathcal{C}\subseteq \mathbb{F}_2^n$. Usually, such a code is specified by its \textit{parity check matrix} $H \in \mathbb{F}_2^{\ell \times n}$ with $\ell \geq n-k$ which is a matrix $H$ such that $\mathcal{C} = \ker (H)$. We will often write $\mathcal{C}_H$ for the code induced by a parity check matrix $H$.
The \textit{distance} of a classical linear code is the minimal Hamming weight of its non-zero elements.

A linear code $\mathcal{C}\subseteq \mathbb{F}_2^n$ is \emph{cyclic} if it is invariant under cyclic shifts of codewords, that is,
\begin{equation}\label{eq:cycliccodecondition}
    v=(a_0,a_1,\dots,a_{n-1})\in \mathcal{C} \implies S_nv=(a_{n-1},a_0,\dots,a_{n-2})\in \mathcal{C}, \text{ for all }v\in \mathbb{F}_2^n
\end{equation}
where $S_n\in \mathbb{F}_2^{n\times n}$ is the cyclic permutation matrix 
\begin{equation}\label{eq:Smatrix}
    S_{\ell} = \begin{pmatrix}
    &&&1\\
    1&& &\\
    &\ddots&&\\
    &&1&
    \end{pmatrix}.
\end{equation}

Cyclic codes can be neatly described using polynomials. Denote by $\mathbb{F}_2[x]$ the ring of univariate polynomials over $\mathbb{F}_2$ and by $\mathbb{F}_2[x]_n\subset \mathbb{F}_2[x]$ the set of polynomials of degree less than $n$. We may identify $\mathbb{F}_2^n\cong \mathbb{F}_2[x]_n\cong R=\mathbb{F}_2[x]/(x^n-1)$ via
\begin{equation}\label{eq:vectorstorpolynomials}
    a=(a_0,\dots,a_{n-1}) \mapsto f_a =\sum_{i=0}^{n-1}a_ix^i.
\end{equation}

Abusing notation, we will write $\mathcal{C}$ for both the codespace as a subspace of $\mathbb{F}_2^{n}$ and $R$. In $R$, the condition \eqref{eq:cycliccodecondition} translates to $f \in \mathcal{C}$ implying  $xf \in \mathcal{C}$ for all $f \in R$. Since $\mathcal{C}$ is linear, this is equivalent to the condition that $\mathcal{C} \subseteq R$ is an ideal,
that is, $\mathcal{C}$ is closed under addition and under multiplication with any element in $R.$ Since $R$ is a principal ideal domain, we can find for any cyclic code $\mathcal{C}$ a unique monic \emph{generator polynomial} $g\in \mathcal{C}$ of minimal degree such that $\mathcal{C}=(g)=\{fg\mid f\in R\}$. Using the Euclidean algorithm, one can see that then there exists a unique polynomial $h$ such that $g h = x^n -1$ and $\mathcal{C} = \lbrace f: fh = 0 \mod x^n -1 \rbrace $. The polynomial $h$ is usually called the \textit{check polynomial} for the code $\mathcal{C}$. Then $h(S_n)$ is a parity check matrix for $\mathcal{C}$, that is, $\mathcal{C}  = \ker (h(S_n))$. 

One can read off important properties of the code from the generator and check polynomials.

\begin{lemma}\label{lem: deleterowsfromparitycheck}
    Let $\mathcal{C}\subseteq \mathbb{F}_2^n$ be a cyclic code with generator polynomial $g$ and check polynomial $h$ such that $\deg (h) = r$. Then, $\text{dim} (\mathcal{C}) = r$. More precisely, any choice of $n-r$ consecutive rows of $h(S_n)$ are linearly independent and span the remaining rows.
\end{lemma}
\begin{proof}
The polynomials $h,\dots ,x^{n - r -1}h$ have pairwise different degrees less than $n$, therefore no linear combination of them is zero. Multiplying with $x^a$ shows the same for $x^a h, \dots , x^{a + n - r -1}h$. Hence, 
using the correspondence~\eqref{eq:vectorstorpolynomials}, we see that any $n-r$ consecutive rows of $h(S_n)$ are linearly independent. The relation $gh = x^n -1$ shows that the span of $h, \dots , x^{n - r -1}h$ contains $x^{n-r} h$. Multiplying again with powers of $x$ shows that in fact all rows are in the span of $h, \dots , x^{n - r -1}h$. This shows the second claim which also implies that $\mathcal{C} = \ker (h(S_n))$ has dimension $r$.
\end{proof}

Let now $h \in \mathbb{F}_2[x]$ be a polynomial of degree $r < n$. Then, $h(S_n)$ defines a parity check matrix and the associated code is a cyclic code. One easily sees that via the identification~\eqref{eq:vectorstorpolynomials}
\begin{equation*}
    h(S_n) v = 0 \iff hf_v = 0 \mod x^n -1.
\end{equation*}
In fact, the inner product of the $i$-th row of $h(S_n)$ with the vector $v$ is exactly the coefficient of $x^i$ of $h f_v$.

\begin{lemma}\label{lem:kergcdlemma}
    It holds that $\ker (\gcd (h, x^n - 1)(S_n)) = \ker (h(S_n))$. In particular,  $h(S_n)$ has non-trivial kernel if and only if $\gcd(h, x^n -1 ) \neq 1$.
\end{lemma}
\begin{proof}
    Clearly, $\ker (\gcd (h, x^n - 1)(S_n)) \subset \ker (h(S_n))$. For the other direction, say $h(S_n)v = 0$. We can find $a,b \in \mathbb{F}_2[x]$ to write $\gcd (h, x^n - 1) = ah + b(x^n - 1)$ using the Euclidean algorithm. Then, 
        $\gcd (h,x^n - 1)(S_n) v = a(S_n)h(S_n)v + b(S_n) (S_n^n - \mathds{1}_n) v = 0$
    since $S_n^n = \mathds{1}_n$.
\end{proof}

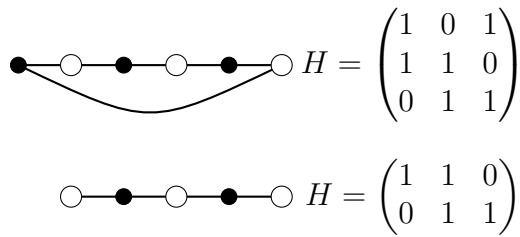
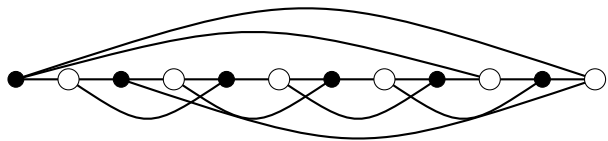
\begin{figure}
     \centering
     \begin{subfigure}[b]{0.4\textwidth}
         \centering
             \begin{tikzpicture}[scale = .7]
        
        \draw[thick] (-1,0) -- (4,0) .. controls (1.5,-1.2) .. cycle;
        \draw[fill = classicalbitcolor] (0,0) circle (.2);
        \draw[fill = classicalbitcolor] (2,0) circle (.2);
        \draw[fill = classicalbitcolor] (4,0) circle (.2);
        
        \draw[fill = classicalcheckcolor] (-1,0) circle (.15);
        \draw[fill = classicalcheckcolor] (1,0) circle (.15);
        \draw[fill = classicalcheckcolor] (3,0) circle (.15);

\node at (6.5,0) {$H = \begin{pmatrix}
    1& 0 & 1\\ 1 & 1 & 0 \\0 & 1 & 1
\end{pmatrix}$};

        \draw[thick] (0,-2.5) -- (4,-2.5);
        \draw[fill = classicalbitcolor] (0,-2.5) circle (.2);
        \draw[fill = classicalbitcolor] (2,-2.5) circle (.2);
        \draw[fill = classicalbitcolor] (4,-2.5) circle (.2);

        \draw[fill = classicalcheckcolor] (1,-2.5) circle (.15);
        \draw[fill = classicalcheckcolor] (3,-2.5) circle (.15);
\node at (6.5,-2.5) {$H = \begin{pmatrix}
    1& 1 & 0\\ 0 & 1 & 1 
\end{pmatrix}$};
        
    \end{tikzpicture}
         \caption{Two possible Tanner graphs of the repetition code.}
         \label{fig:y equals x}
     \end{subfigure}
     \hfill
     \begin{subfigure}[b]{0.5\textwidth}
         \centering
         \begin{tikzpicture}[scale = .7]
    \draw[classicalcheckcolor, thick] (1,0) -- (12,0);
    \foreach \i in {2,4,6,8} (
    \draw[classicalcheckcolor,thick] (\i,0)  .. controls (\i + 1.5,-1) ..(\i + 3,0);
    )
    
    \draw[classicalcheckcolor,thick] (1,0)  .. controls (5.5,1.2) ..(10,0);
    \draw[classicalcheckcolor,thick] (1,0)  .. controls (6.5,1.8) ..(12,0);
    \draw[classicalcheckcolor,thick] (3,0)  .. controls (7.5,-1.5) ..(12,0);
    
        \foreach \i in {1,2,3,4,5,6} (
            \draw[fill = classicalcheckcolor] (2*\i-1,0) circle (.15);
         )
         \foreach \i in {1,2,3,4,5,6} (
            \draw[fill = classicalbitcolor] (2*\i,0) circle (.2);
         )
        
    \end{tikzpicture}   
         \caption{The Tanner graph of the cyclic code associated with $1 + x + x^2$}
         \label{fig:five over x}
     \end{subfigure}
        \caption{Examples for Tanner graphs. The variable vertices are white, the check vertices black. In (a), we display two possible Tanner graphs for the repetition code that encodes 1 logical bit into 3 physical bits via $0 \mapsto 000, 1 \mapsto 111$. The first one is the cyclic code with parity check matrix $h(S_3)$ where $h = 1 + x$. For the second one, we discard the first (redundant) check. In (b), we display the Tanner graph of the cyclic code associated with $h(S_6)$ where $h = 1 + x  + x^2$.}
        \label{fig:tanner graphs}
\end{figure}

Often, it is helpful to visualize a code $\mathcal{C}$ by its \textit{Tanner graph}. Given a code $\mathcal{C} \subseteq \mathbb{F}_2^n$ with parity check $H \in \mathbb{F}_2^{\ell \times n}$, the Tanner graph is a graph $G = (V \cup C, E)$ whose vertex set consists of \textit{variable nodes} $V = \lbrace v_1, \dots, v_n \rbrace $ and \textit{check nodes} $C = \lbrace c_1, \dots, c_{\ell} \rbrace$. The edges mark which check is acting on which variable, that is, $E = \left\lbrace \lbrace c_i, v_j \rbrace : H_{i,j} \neq 0 \right\rbrace .$
Note that the Tanner graph depends not only on the code space but on the particular parity check matrix one uses to construct the code. Examples of Tanner graphs are shown in Figure~\ref{fig:tanner graphs}.

Given a classical code $\mathcal{C}$ with associated parity check matrix $H$, its \textit{transpose code} $\mathcal{C}^T$ is defined to be the code with parity check matrix $H^T$, that is, the code space of $\mathcal{C}^T$ is $\ker (H^T)$. The Tanner graph of $\mathcal{C}^T$ arises from the Tanner graph of $\mathcal{C}$ by switching the roles of variable and check nodes. For an $[n,k,d]$ code $\mathcal{C}$, we write $[n^T, k^T, d^T]$ for the parameters of its transpose code $\mathcal{C}^T$. We record one more useful fact.
\begin{lemma}\label{lem: transpose of cyclic code}
    Let $\mathcal{C}$ be a $[n,k,d]$ cyclic code with parity check matrix $h(S_n)$ for some polynomial $h(x)$. Then, $\mathcal{C}^T = \left\lbrace J_n v: v \in \mathcal{C}\right\rbrace $
    where $J_n$ is the $n\times n$ matrix with ones on the anti-diagonal. In particular, $\mathcal{C}^T$ is also a $[n,k,d]$-code. 
\end{lemma}
\begin{proof}
Follows from $S_n^T = J_n S_n J_n$.
\end{proof}

\subsection{CSS codes}
In this section, we will fix some notation regarding quantum stabilizer codes and in particular CSS codes~\cite{PhysRevA.54.1098,doi:10.1098/rspa.1996.0136}. A quantum code encoding $k$ \textit{logical qubits} into $n$ \textit{physical qubits} is a $2^k$-dimensional space $\mathcal{Q} \subseteq (\mathbb{C}^2)^{\otimes n}$. Denote by $X$, $Y$, and $Z$ the Pauli matrices and by $\mathcal{P}_n$ the Pauli group acting on $n$-qubits. Given commuting Pauli operators $S_1, \dots, S_{\ell}$ such that no product of $S_i$'s yields $- \mathds{1}$, the associated \textit{stabilizer code} is the space of all $\ket{\psi} \in (\mathbb{C}^2)^{\otimes n}$ such that $S_i \ket{\psi} = \ket{\psi}$ for all $i = 1, \dots, \ell$. The operators $S_i$ are called the \textit{stabilizer generators}, the group $\mathcal{S} = \langle S_1, \dots, S_{\ell} \rangle$ generated by them is called the \textit{stabilizer group}. It turns out that if exactly $m$ of the $\ell$ stabilizer generators are independent then this stabilizer code encodes $n - m$ logical qubits. 

It is often useful to write the stabilizer generators in the form of a \textit{check matrix} $H = (H_X | H_Z)$. This matrix has $2n$ columns and $\ell$ rows. The stabilizer generator $S_i$ has the form $\bigotimes_{j = 1, \dots, n} X^{H_{i,j}} Z^{H_{i,n + j}}$.

A stabilizer code is a \textit{CSS code} whenever each of the stabilizer generators is either a product of only identities and $X$ operators or a product of only identities and $Z$ operators. For a CSS code, the matrix $H$ is of the form 
\begin{equation*}
   \left(
\begin{array}{c|c}
H_X & 0 \\
0 & H_Z \\
\end{array}
\right).
\end{equation*}

We call $H_X$ and $H_Z$ the $X$-check and $Z$-check matrices, respectively. The condition that all stabilizer generators have to commute translates to the condition that $H_X H_Z^T = 0$.

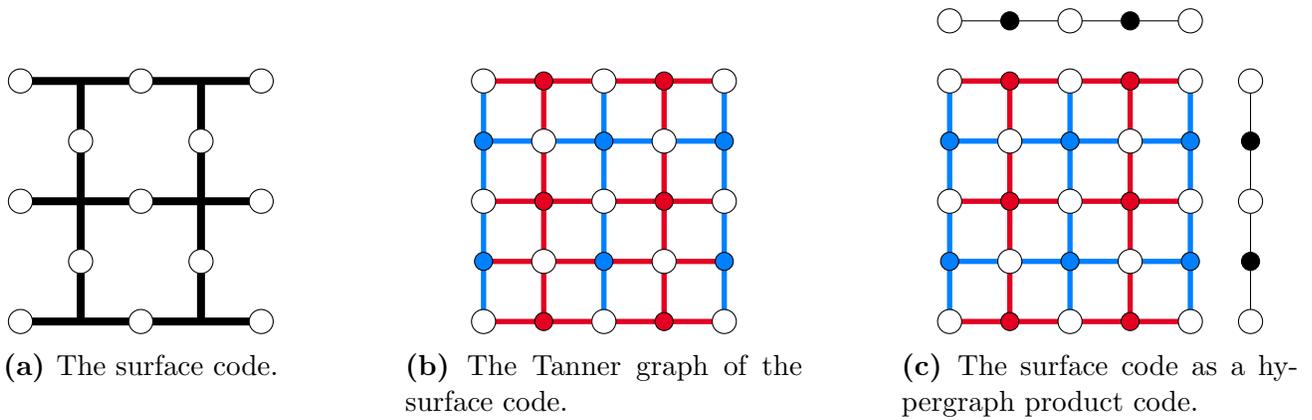
\begin{figure}
     \centering
     \begin{subfigure}[t]{0.25\textwidth}
         \centering
    \begin{tikzpicture}[scale = .8]
    \draw[line width=3pt] (1,0) -- (1,4);
    \draw[line width=3pt] (3,0) -- (3,4);

 \draw[line width=3pt] (0,0) -- (4,0);
 \draw[line width=3pt] (0,2) -- (4,2);
 \draw[line width=3pt] (0,4) -- (4,4);

        \draw[fill = qubitcolor] (0,0) circle (.2);
        \draw[fill = qubitcolor] (2,0) circle (.2);
        \draw[fill = qubitcolor] (4,0) circle (.2);

        \draw[fill = qubitcolor] (1,1) circle (.2);
        \draw[fill = qubitcolor] (3,1) circle (.2);

        \draw[fill = qubitcolor] (0,2) circle (.2);
        \draw[fill = qubitcolor] (2,2) circle (.2);
        \draw[fill = qubitcolor] (4,2) circle (.2);

        \draw[fill = qubitcolor] (1,3) circle (.2);
        \draw[fill = qubitcolor] (3,3) circle (.2);

        \draw[fill = qubitcolor] (0,4) circle (.2);
        \draw[fill = qubitcolor] (2,4) circle (.2);
        \draw[fill = qubitcolor] (4,4) circle (.2);
        \end{tikzpicture}
         \caption{The surface code.}\label{subfig:surfacecode}
     \end{subfigure}
     \hfill
     \begin{subfigure}[t]{0.3\textwidth}
         \centering
\begin{tikzpicture}[scale = .8]
 \draw[line width=2pt,zcheckcolor] (0,0) -- (0,4);
 \draw[line width=2pt,zcheckcolor] (2,0) -- (2,4);
 \draw[line width=2pt,zcheckcolor] (4,0) -- (4,4);
\draw[line width=2pt,xcheckcolor] (1,0) -- (1,4);
 \draw[line width=2pt,xcheckcolor] (3,0) -- (3,4);
\draw[line width=2pt,xcheckcolor] (0,0) -- (4,0);
\draw[line width=2pt,xcheckcolor] (0,2) -- (4,2);
\draw[line width=2pt,xcheckcolor] (0,4) -- (4,4);
        \draw[fill = qubitcolor] (0,0) circle (.2);
        \draw[fill = qubitcolor] (2,0) circle (.2);
        \draw[fill = qubitcolor] (4,0) circle (.2);
        \draw[fill = xcheckcolor] (1,0) circle (.15);
        \draw[fill = xcheckcolor] (3,0) circle (.15);

\draw[line width=2pt,xcheckcolor] (0,1) -- (4,1);

        \draw[fill = zcheckcolor] (0,1) circle (.15);
        \draw[fill = zcheckcolor] (2,1) circle (.15);
        \draw[fill = zcheckcolor] (4,1) circle (.15);

        \draw[fill = qubitcolor] (1,1) circle (.2);
        \draw[fill = qubitcolor] (3,1) circle (.2);

        \draw[fill = qubitcolor] (0,2) circle (.2);
        \draw[fill = qubitcolor] (2,2) circle (.2);
        \draw[fill = qubitcolor] (4,2) circle (.2);

        \draw[fill = xcheckcolor] (1,2) circle (.15);
        \draw[fill = xcheckcolor] (3,2) circle (.15);

\draw[line width=2pt,zcheckcolor] (0,3) -- (4,3);

        \draw[fill = zcheckcolor] (0,3) circle (.15);
        \draw[fill = zcheckcolor] (2,3) circle (.15);
        \draw[fill = zcheckcolor] (4,3) circle (.15);

        \draw[fill = qubitcolor] (1,3) circle (.2);
        \draw[fill = qubitcolor] (3,3) circle (.2);

        \draw[fill = qubitcolor] (0,4) circle (.2);
        \draw[fill = qubitcolor] (2,4) circle (.2);
        \draw[fill = qubitcolor] (4,4) circle (.2);

        \draw[fill = xcheckcolor] (1,4) circle (.15);
        \draw[fill = xcheckcolor] (3,4) circle (.15);
    \end{tikzpicture}
         \caption{The Tanner graph of the surface code.}
         \label{fig:tannergraphsurfacecode}
     \end{subfigure}
     \hfill
    \begin{subfigure}[t]{0.3\textwidth}
    \centering
               \begin{tikzpicture}[scale = .8]

 \draw[line width=2pt,zcheckcolor] (0,0) -- (0,4);
 \draw[line width=2pt,zcheckcolor] (2,0) -- (2,4);
 \draw[line width=2pt,zcheckcolor] (4,0) -- (4,4);

\draw[line width=2pt,xcheckcolor] (1,0) -- (1,4);
 \draw[line width=2pt,xcheckcolor] (3,0) -- (3,4);

\draw[line width=2pt,xcheckcolor] (0,0) -- (4,0);
\draw[line width=2pt,xcheckcolor] (0,2) -- (4,2);
\draw[line width=2pt,xcheckcolor] (0,4) -- (4,4);

        \draw[fill = qubitcolor] (0,0) circle (.2);
        \draw[fill = qubitcolor] (2,0) circle (.2);
        \draw[fill = qubitcolor] (4,0) circle (.2);

        \draw[fill = xcheckcolor] (1,0) circle (.15);
        \draw[fill = xcheckcolor] (3,0) circle (.15);

\draw[line width=2pt,zcheckcolor] (0,1) -- (4,1);

        \draw[fill = zcheckcolor] (0,1) circle (.15);
        \draw[fill = zcheckcolor] (2,1) circle (.15);
        \draw[fill = zcheckcolor] (4,1) circle (.15);

        \draw[fill = qubitcolor] (1,1) circle (.2);
        \draw[fill = qubitcolor] (3,1) circle (.2);

        \draw[fill = qubitcolor] (0,2) circle (.2);
        \draw[fill = qubitcolor] (2,2) circle (.2);
        \draw[fill = qubitcolor] (4,2) circle (.2);

        \draw[fill = xcheckcolor] (1,2) circle (.15);
        \draw[fill = xcheckcolor] (3,2) circle (.15);

\draw[line width=2pt,zcheckcolor] (0,3) -- (4,3);

        \draw[fill = zcheckcolor] (0,3) circle (.15);
        \draw[fill = zcheckcolor] (2,3) circle (.15);
        \draw[fill = zcheckcolor] (4,3) circle (.15);

        \draw[fill = qubitcolor] (1,3) circle (.2);
        \draw[fill = qubitcolor] (3,3) circle (.2);

        \draw[fill = qubitcolor] (0,4) circle (.2);
        \draw[fill = qubitcolor] (2,4) circle (.2);
        \draw[fill = qubitcolor] (4,4) circle (.2);

        \draw[fill = xcheckcolor] (1,4) circle (.15);
        \draw[fill = xcheckcolor] (3,4) circle (.15);

        \draw (0,5) -- (4,5);
        \draw[fill = classicalbitcolor] (0,5) circle (.2);
        \draw[fill = classicalbitcolor] (2,5) circle (.2);
        \draw[fill = classicalbitcolor] (4,5) circle (.2);

        \draw[fill = classicalcheckcolor] (1,5) circle (.15);
        \draw[fill = classicalcheckcolor] (3,5) circle (.15);

 \draw (5,0) -- (5,4);
        \draw[fill = classicalbitcolor] (5,0) circle (.2);
        \draw[fill = classicalbitcolor] (5,2) circle (.2);
        \draw[fill = classicalbitcolor] (5,4) circle (.2);

        \draw[fill = classicalcheckcolor] (5,1) circle (.15);
        \draw[fill = classicalcheckcolor] (5,3) circle (.15);
    \end{tikzpicture}
    \caption{The surface code as a hypergraph product code.}\label{fig:surface code as hypergraph product}
    \end{subfigure}
        \caption{For the surface code, the qubits are placed on the edges of a lattice with two \textit{rough} and two \textit{smooth} boundaries as shown in (a). For each plaquette of the lattice $P$, there is a stabilizer $\prod_{i \in P} Z_i$. For each vertex $v$ of the lattice, there is a stabilizer $\prod_{v \in i} X_i$. In (b), we show the corresponding Tanner graph with $X$-check vertices in red and $Z$-check vertices in blue. Finally, in (c), we see how the surface code arises as a hypergraph product $\mathcal{C} \hgproduct \mathcal{C}$ where $\mathcal{C} $ is just the repetition code as in Figure~\ref{fig:tanner graphs}.}\label{fig:surface code}
\end{figure}

CSS codes can, similar to classical linear codes, be visualized using Tanner graphs. For a CSS code with $X$- and $Z$-check matrices $H_X$ and $H_Z$, the Tanner graph $G = (V\cup C^X \cup C^Z, E)$ has a node $v_j$ for each qubit, a node $c^X_i$ for each $X$-check and a node $c^Z_i$ for each $Z$-check. Again, we draw an edge between $c^X_i$ (resp. $c^Z_i$) and $v_j$ if the corresponding $X$-check (resp. $Z$-check) acts non-trivial on the qubit $j$, that is, if $(H_X)_{ij} =1$ resp. $(H_Z)_{ij} =1$. An example of a Tanner graph can be found in Figure~\ref{fig:tannergraphsurfacecode}.

The \textit{logical operators} of a stabilizer code $\mathcal{Q}$ specified by stabilizer generators $S_1, \dots, S_{\ell}$ are the elements of $\mathcal{P}_n$ that commute with each of the $S_i$ but are not in the group generated by the $S_i$. The \textit{distance} of a stabilizer code is the minimum Hamming weight of a non-trivial logical operator. For a code $\mathcal{Q}$ encoding $k$ logical qubits into $n$ physical qubits that has distance $d$ we say that $\mathcal{Q}$ is an $[[n,k,d]]$ code.

\subsection{Hypergraph product codes}

The \textit{hypergraph product code} construction introduced in~\cite{Tillich_2014} is a systematic way of constructing a quantum CSS code from a pair of classical codes. Let $\mathcal{C}_1$ and $\mathcal{C}_2$ be classical linear codes with associated parity check matrices $H_1 \in \mathbb{F}_2^{\ell_1 \times n_1}$ and $H_2\in \mathbb{F}_2^{\ell_2 \times n_2}$. Define 
\begin{equation}\label{eq:tensorproductparitycheck}
    H_X = (H_1 \otimes \mathds{1}_{n_2} | \mathds{1}_{\ell_1}\otimes H_2^T),\;\; H_Z = (\mathds{1}_{n_1} \otimes H_2 | H_1^T \otimes \mathds{1}_{\ell_2}).
\end{equation}
Then $H_X H_Z^T = 0$, that is, $H_X$ and $H_Z$ define $X$- and $Z$-checks for a CSS code $\mathcal{Q} = \mathcal{C}_1 \hgproduct \mathcal{C}_2$. The code $\mathcal{Q}$ is called the hypergraph product of $\mathcal{C}_1$ and $\mathcal{C}_2$. 

The hypergraph product construction can be nicely understood in terms of Tanner graphs. Denote by $G_1 = (V_1 \cup C_1, E_1)$ and $G_2 = (V_2 \cup C_2, E_2)$ denote the Tanner graphs of $\mathcal{C}_1$ and $\mathcal{C}_2$, respectively. Let $G_1 \times G_2$ be the cartesian product of the graphs $G_1$ and $G_2$. Letting 
\begin{equation*}
   V = V_1 \times V_2 \cup C_1 \times C_2, \;\; C_X = C_1 \times V_2, \;\; C_Z = V_1 \times C_2,
\end{equation*}

 this turns out to define the Tanner graph of a well-defined CSS code which is exactly the hypergraph product code $\mathcal{C}_1 \hgproduct \mathcal{C}_2$. For example, the surface code is a hypergraph product of two copies of the repetition codes, for a visualization of this see Figure~\ref{fig:surface code as hypergraph product}. Another example of the hypergraph product of two cyclic codes can be seen in Figure~\ref{fig:tannergraphtensorproduct}.

The parameters of the hypergraph product of two classical codes only depend on the parameters of the classical codes.
\begin{theorem}[\cite{Tillich_2014}] \label{lem: tensorproductparameters}
    Let $\mathcal{C}_1$ and $\mathcal{C}_2$ be classical linear codes with parameters $[n_1,k_1,d_1]$ and $[n_2,k_2,d_2]$, respectively, induced by parity check matrices $H_1 \in \mathbb{F}_2^{\ell_1 \times n_1}$ and $H_2 \in \mathbb{F}_2^{\ell_2 \times n_2}$. Then, $\mathcal{C}_1 \hgproduct \mathcal{C}_2$ is a quantum error-correcting code with parameters 
    \begin{equation*}
        [[n_1n_2 + \ell_1 \ell_2, k_1k_2 + k_1^Tk_2^T, \min(d_1,d_2,d_1^T,d_2^T)]].
    \end{equation*}
\end{theorem}

If any of the codes $\mathcal{C}_1, \mathcal{C}_2, \mathcal{C}_1^T$ or $\mathcal{C}_2^T$ are trivial, we take by convention the distance $d_1, d_2,d_1^T$ resp. $d_2^T$ of this code to be  $\infty$. 
Using Lemma~\ref{lem: transpose of cyclic code}, we can see that if $\mathcal{C}_1$ and $\mathcal{C}_2$ are, respectively, cyclic $[n_1,k_1,d_1]$ and $[n_2,k_2,d_2]$ codes specified by parity check matrices $h_1(S_{n_1})$ and $h_2(S_{n_2})$, then $\mathcal{C}_1 \hgproduct \mathcal{C}_2$ has parameters $[2n_1n_2, 2k_1k_2, \min(d_1,d_2)]].$

\subsection{Bivariate bicycle codes}\label{sec:bivariatebicyclecodes}

The bivariate bicycle codes introduced in~\cite{PhysRevA.88.012311} are a natural generalization of hypergraph products of pairs of cyclic codes. 
This generalization comes naturally when rephrasing the definition of hypergraph product codes in the following way: Consider two classical cyclic codes $\mathcal{C}_1$ and $\mathcal{C}_2$ given by parity check matrices $h_1(S_{\ell})$ and $h_2(S_m)$. One can construct the $X$- and $Z$-check matrices of $\mathcal{C}_1 \hgproduct \mathcal{C}_2^T$ from $h_1(S_{\ell})$ and $h_2(S_m)$ using the formula~\eqref{eq:tensorproductparitycheck}\footnote{Here, we use $\mathcal{C}_2^T$ instead of $\mathcal{C}_2$ to match notation with prior work on bivariate bicycle codes~\cite{PhysRevA.88.012311,bravyi2023highthreshold}.}. An equivalent way of constructing the check matrices is by defining 
\begin{equation*}
        x = S_{\ell} \otimes \mathds{1}_m , \;\; y = \mathds{1}_{\ell} \otimes S_m.
\end{equation*}

With that, we see that the parity check matrices of the hypergraph product of $\mathcal{C}_1$ and $\mathcal{C}_2^T$ arise as 
\begin{equation}\label{eq:checkbivariatetensorproduct}
    H_X = (h_1(x)|h_2(y)), H_Z = (h_1(y)^T|h_2(x)^T).
\end{equation}

Now, bivariate bicyclic codes are the natural generalization of codes with check matrices as in~\eqref{eq:checkbivariatetensorproduct} where we allow bivariate polynomials instead of single variable polynomials: Letting $A(x,y)$ and $B(x,y)$ be bivariate polynomials, we can define $X$- and $Z$-check matrices 
\begin{equation}\label{eq:checkbivariatepolynomials}
    H_X = (A(x,y)|B(x,y)), \;\; H_Z = (B(x,y)^T|A(x,y)^T).
\end{equation}

We notice from the visualization in Figure~\ref{fig:bivariatebicyclestabs} that $\max (\deg (A(x,y)),\deg (B(x,y)) )$ quantifies how far a syndrome qubit is from the data qubits it checks, that is, how non-local the stabilizer checks are. 

\section{Towards bivariate bicycle codes with open boundary conditions}\label{sec: towardsopenboundary}
We saw that bivariate bicycle codes have the nice property that we can tune how local their checks are with respect to a 2D lattice by imposing bounds on the degree of the polynomials $A(x,y)$ and $B(x,y)$. This locality is with respect to a 2D lattice with periodic boundary conditions. For practical purposes, we actually desire local checks with respect to a lattice with open boundary conditions. In this section, we propose ways of transforming bivariate bicycle codes to codes with similar locality properties on 2D lattices with open boundary conditions. In particular, we show that this is always possible for hypergraph products of cyclic codes, that is, for bivariate bicycle codes induced by single variable polynomials $A(x)$ and $B(y)$. We then discuss to what extent this is possible for more general bivariate bicycle codes and give examples of bivariate bicycle codes that are not hypergraph products.

\begin{definition}
    Let $\mathcal{Q}$ be a quantum stabilizer code associated with stabilizer generators $S_1, \dots, S_{\ell}$ and set of physical qubits $\Lambda = \lbrace 1, \dots, n \rbrace$. A code $\tilde{\mathcal{Q}}$ is a \textit{pruned code derived from} $\mathcal{Q}$ if there are subsets $\tilde{\Lambda} \subset \Lambda$ and $I \subset \lbrace 1, \dots, l \rbrace$ such that $\tilde{\mathcal{Q}}$ is a quantum error-correcting code defined on qubits $\tilde{\Lambda}$ associated with stabilizer generators $S_i|_{\tilde{\Lambda}}, i \in I$.  
\end{definition}

We will often refer to $\tilde{\mathcal{Q}}$ as a \textit{pruned version} of $\mathcal{Q}$. Sometimes, we will simply say that we pruned $\mathcal{Q}$. Note that not for all choices of $\tilde{\Lambda}$ and $I$ this gives a valid stabilizer code since the remaining stabilizers might not commute. For a CSS code $\mathcal{Q}$ with $X$- and $Z$-check matrices $H_X$ and $H_Z$, pruning it corresponds to  deleting the same set of columns from $H_X$ and $H_Z$, and deleting some rows on each giving rise to matrices $\tilde{H}_X$ and $\tilde{H}_Z$ such that $\tilde{H}_X\tilde{H}_Z^T = 0$ holds. In the following, fix $\ell$ and $m$, and let $A(x)$ and $B(y)$ be univariate polynomials of degree $r_A<l$ and $r_B<m$, respectively, and write $A_{red}$ for the matrix arising by deleting the first $r_A$ rows from $A(S_{\ell})$ and $B_{red}$ for the matrix arising by deleting the last $r_B$ columns from $B(S_m)$.

\begin{lemma}\label{lem: subcodewithoutperiodicoftensorproduct}
    In the set-up as before, define $\mathcal{Q}$ as the bivariate bicycle code associated with the polynomials $A(x)$ and $B(y)$. 
    Then, the hypergraph product code $\mathcal{C}_{A_{red}} \hgproduct \mathcal{C}_{B_{red}}^T$ is a pruned version of $\mathcal{Q}$ where $\mathcal{C}_{A_{red}}$ is the classical cyclic code with parity check matrix $A_{red}$ and similar for $\mathcal{C}_{B_{red}}$.
\end{lemma}

\begin{figure}
    \begin{subfigure}[t]{0.45\textwidth}
     \begin{tikzpicture}[scale = .4]

    \begin{scope}[yshift = -1cm]
        \draw[classicalcheckcolor, thick] (1,0) -- (12,0);
    \foreach \i in {2,4,6,8} (
    \draw[classicalcheckcolor,thick] (\i,0)  .. controls (\i + 1.5,-1) ..(\i + 3,0);
    )
    
    \draw[classicalcheckcolor,thick] (1,0)  .. controls (5.5,1.2) ..(10,0);
    \draw[classicalcheckcolor,thick] (1,0)  .. controls (6.5,1.8) ..(12,0);
    \draw[classicalcheckcolor,thick] (3,0)  .. controls (7.5,-1.5) ..(12,0);
    
        \foreach \i in {1,2,3,4,5,6} (
            \draw[fill = classicalcheckcolor] (2*\i-1,0) circle (.15);
         )
         \foreach \i in {1,2,3,4,5,6} (
            \draw[fill = classicalbitcolor] (2*\i,0) circle (.2);
         )
    \end{scope}

    \draw[classicalcheckcolor, thick] (-1,1) -- (-1,12);
    \foreach \i in {2,4,6,8} (
    \draw[classicalcheckcolor,thick] (-1,\i)  .. controls (-2,\i + 1.5) ..(-1,\i + 3);
    )
    
    \draw[classicalcheckcolor,thick] (-1,1)  .. controls (0.2,5.5) ..(-1,10);
    \draw[classicalcheckcolor,thick] (-1,1)  .. controls (.7,6.5) ..(-1,12);
    \draw[classicalcheckcolor,thick] (-1,3)  .. controls (-2.5,7.5) ..(-1,12);
    
        \foreach \i in {1,2,3,4,5,6} (
            \draw[fill = classicalbitcolor] (-1,2*\i-1) circle (.2);
         )
         \foreach \i in {1,2,3,4,5,6} (
            \draw[fill = classicalcheckcolor] (-1,2*\i) circle (.15);
         )


\foreach \x in {1,3,5,7,9,11}(
\draw[thick,color = black!50!white] (1,\x) -- (12.5,\x);
)

\foreach \x in {1,3,5,7,9,11}(
\draw[thick,color = black!50!white] (\x,1) -- (\x,12.5);
)

     \draw[color = zcheckcolor,very thick] (8,5) -- (8,7);
\draw[color = zcheckcolor,very thick] (7,6) -- (9,6);
\draw[color = zcheckcolor,very thick] (8,6) .. controls (7.3,7.5) ..(8,9);
\draw[color = zcheckcolor,very thick] (8,6) .. controls (9.5,6.7) ..(11,6);

\draw[color = xcheckcolor,very thick] (5,8) -- (5,10);
\draw[color = xcheckcolor,very thick] (4,9) -- (6,9);
\draw[color = xcheckcolor,very thick] (5,9) .. controls (3.5,9.7) ..(2,9);
\draw[color = xcheckcolor,very thick] (5,9) .. controls (4.3,7.5) ..(5,6);

    \foreach \x in {1,3,5,7,9,11} (
    \foreach \y in {1,3,5,7,9,11} (
    \draw[fill = xcheckcolor] (\x,\y) circle (.15);
    ))
    \foreach \x in {1,3,5,7,9,11} (
    \foreach \y in {1,3,5,7,9,11} (
    \draw[fill = zcheckcolor] (\x+1,\y+1) circle (.15);
    ))

    \foreach \x in {1,3,5,7,9,11} (
    \foreach \y in {1,3,5,7,9,11} (
    \draw[fill = qubitcolor] (\x+1,\y) circle (.2);
    ))

    \foreach \x in {1,3,5,7,9,11} (
    \foreach \y in {1,3,5,7,9,11} (
    \draw[fill = qubitcolor] (\x,\y+1) circle (.2);
    ))

    \end{tikzpicture}   
        \caption{The Tanner graph of the bivariate bicycle code associated with $A(x) = 1 + x + x^2$ and $B(y) = 1 + y + y^2$.}
    \label{fig:tannergraphtensorproduct}
    \end{subfigure}
    \hfill
    \begin{subfigure}[t]{0.45\textwidth}
     \begin{tikzpicture}[scale = .4]
    \begin{scope}[yshift = -1cm]
    \draw[color =green!80!black, fill=green!80!black,opacity = .2] (.5,0.5) -- (1.5,0.5) -- (1.5,-.5) -- (.5,-.5) -- cycle;
    \draw[color =green!80!black, fill=green!80!black,opacity = .2] (2.5,0.5) -- (3.5,0.5) -- (3.5,-.5) -- (2.5,-.5) -- cycle;
        \draw[classicalcheckcolor, thick] (1,0) -- (12,0);
    \foreach \i in {2,4,6,8} (
    \draw[classicalcheckcolor,thick] (\i,0)  .. controls (\i + 1.5,-1) ..(\i + 3,0);
    )
    
    \draw[classicalcheckcolor,thick] (1,0)  .. controls (5.5,1.2) ..(10,0);
    \draw[classicalcheckcolor,thick] (1,0)  .. controls (6.5,1.8) ..(12,0);
    \draw[classicalcheckcolor,thick] (3,0)  .. controls (7.5,-1.5) ..(12,0);
    
        \foreach \i in {1,2,3,4,5,6} (
            \draw[fill = classicalcheckcolor] (2*\i-1,0) circle (.15);
         )
         \foreach \i in {1,2,3,4,5,6} (
            \draw[fill = classicalbitcolor] (2*\i,0) circle (.2);
         )
    \end{scope}

    \draw[color =green!80!black, fill=green!80!black,opacity = .2] (-.5,9.5) -- (-.5,10.5) -- (-1.5,10.5) -- (-1.5,9.5) -- cycle;
    \draw[color =green!80!black, fill=green!80!black,opacity = .2] (-.5,11.5) -- (-.5,12.5) -- (-1.5,12.5) -- (-1.5,11.5) -- cycle;
    
    \draw[classicalcheckcolor, thick] (-1,1) -- (-1,12);
    \foreach \i in {2,4,6,8} (
    \draw[classicalcheckcolor,thick] (-1,\i)  .. controls (-2,\i + 1.5) ..(-1,\i + 3);
    )
    
    \draw[classicalcheckcolor,thick] (-1,1)  .. controls (0.2,5.5) ..(-1,10);
    \draw[classicalcheckcolor,thick] (-1,1)  .. controls (.7,6.5) ..(-1,12);
    \draw[classicalcheckcolor,thick] (-1,3)  .. controls (-2.5,7.5) ..(-1,12);
    
        \foreach \i in {1,2,3,4,5,6} (
            \draw[fill = classicalbitcolor] (-1,2*\i-1) circle (.2);
         )
         \foreach \i in {1,2,3,4,5,6} (
            \draw[fill = classicalcheckcolor] (-1,2*\i) circle (.15);
         )


    \fill[green!80!black,opacity = .2] (0.5,.5) -- (.5,12.5) -- (12.5,12.5) -- (12.5,9.5) -- (3.5,9.5) -- (3.5,0.5)  -- cycle;

    \fill[white] (1.5,.75) -- (1.5,9.5) -- (2.5,9.5) -- (2.5,.75) -- cycle;%
    \fill[white] (3.5,10.5) -- (3.5,11.5) -- (12.25,11.5) -- (12.25,10.5) -- cycle;

    \fill[color = white]  (1.5,11.5) -- (1.5,10.5) -- (2.5,10.5) -- (2.5,11.5) -- cycle;

\foreach \x in {1,3,5,7,9,11}(
\draw[thick,color = black!50!white] (1,\x) -- (12.5,\x);
)

\foreach \x in {1,3,5,7,9,11}(
\draw[thick,color = black!50!white] (\x,1) -- (\x,12.5);
)

    \foreach \x in {1,3,5,7,9,11} (
    \foreach \y in {1,3,5,7,9,11} (
    \draw[fill = xcheckcolor] (\x,\y) circle (.15);
    ))
    \foreach \x in {1,3,5,7,9,11} (
    \foreach \y in {1,3,5,7,9,11} (
    \draw[fill = zcheckcolor] (\x+1,\y+1) circle (.15);
    ))

    \foreach \x in {1,3,5,7,9,11} (
    \foreach \y in {1,3,5,7,9,11} (
    \draw[fill = qubitcolor] (\x+1,\y) circle (.2);
    ))

    \foreach \x in {1,3,5,7,9,11} (
    \foreach \y in {1,3,5,7,9,11} (
    \draw[fill = qubitcolor] (\x,\y+1) circle (.2);
    ))

    \end{tikzpicture}   
        \caption{The pruned version of it that does not utilize the periodic boundary conditions. The qubits and stabilizers in the green-shaded area are removed.}
    \label{fig:cuttingsubfigure}
    \end{subfigure}
    \caption{The bivariate bicycle code $\mathcal{Q}$ associated with $A(x) = 1 + x + x^2$ and $B(y) = 1 + y + y^2$ is the hypergraph product of the classical cyclic code associated with $1 + x + x^2$ and its transpose. We depict the edges in the Tanner graph for two specific stabilizer generators in (a). By disregarding two redundant checks in the classical codes as in (b), the remaining hypergraph product code is the pruned version of $\mathcal{Q}$ from Lemma~\ref{lem: subcodewithoutperiodicoftensorproduct} which has no edges in the Tanner graph going over the periodic boundary.}
    \label{fig:cuttingcode}
\end{figure}
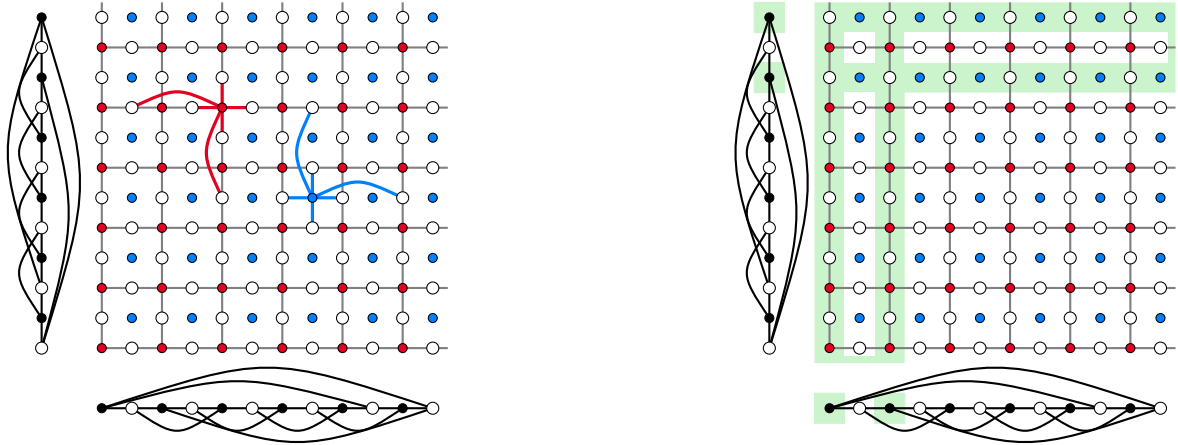

\begin{proof}
Effectively, we just removed columns and rows of the $X$- and $Z$-check matrices. For a visual example, see Figure~\ref{fig:cuttingsubfigure}.
\end{proof}

Using the way of putting syndrome and data qubits as in Figure~\ref{fig:cuttingcode}, we see that the hypergraph product code $\mathcal{C}_{A_{red}} \hgproduct \mathcal{C}_{B_{red}}^T$ in Lemma~\ref{lem: subcodewithoutperiodicoftensorproduct} is by construction as local as the bivariate bicycle code from which we derived it by pruning but does not utilize the periodic boundary conditions. 

It is interesting to compare the parameters of the two codes in Lemma~\ref{lem: subcodewithoutperiodicoftensorproduct}. For that, first assume that $A(x)|(x^{\ell} - 1)$ and $B(y) | (y^m - 1)$. 
Combining Lemma~\ref{lem: deleterowsfromparitycheck} and Lemma~\ref{lem: tensorproductparameters}, we know that the classical cyclic codes $\mathcal{C}_{A(S_{\ell})}$ and $\mathcal{C}_{B(S_m)}$ have parameters $[\ell, r_A, d_A]$ and $[m, r_B, d_B]$ for some $d_A$ and $d_B$, respectively, and that $\mathcal{Q}$ has parameters $[[2\ell m, 2r_Ar_B, \min (d_A,d_B)]]$. 
Lemma~\ref{lem: deleterowsfromparitycheck} also implies that $\mathcal{C}_{A_{red}}$ and $\mathcal{C}_{B_{red}}^T$ have the same parameters as $\mathcal{C}_{A(S_{\ell})}$ and $\mathcal{C}_{B(S_m)}$, respectively, and that $\mathcal{C}_{A_{red}}^T$ and $\mathcal{C}_{B_{red}}$ are trivial. Plugging this into Lemma~\ref{lem: tensorproductparameters}, we obtain the following. 

\begin{proposition}\label{prop: cutting}
    Given $A(x)|(x^{\ell} - 1)$ and $B(y) |(y^m - 1)$, the code $\mathcal{C}_{A_{red}} \hgproduct \mathcal{C}_{B_{red}}^T$ in Lemma~\ref{lem: subcodewithoutperiodicoftensorproduct} has parameters
\begin{equation*}
    [[\ell m + (\ell-r_A)(m-r_B), r_Ar_B, \min (d_A,d_B)]].
\end{equation*}
\end{proposition}

\begin{figure}[hbt!]
     \centering
     \begin{subfigure}[t]{0.25\textwidth}
         \centering
\begin{tikzpicture}[scale = .8,rotate=270]
          \begin{scope}[xshift = -28]
    \draw[fill = zcheckcolor] (4,2) rectangle ++(1,1);
\end{scope}

        \foreach \x in {2,3,4,5,6,7} (
            \draw[very thick, color = black!60!white] (\x,0) -- (\x, 6);
        )
        \foreach \x in {0,1,2,3,4,5} (
            \draw[very thick, color = black!60!white] (1,\x) -- (7,\x);
        )

        \foreach \x in {1,2,3,4,5,6 } (
            \foreach \y in {0,1,2,3,4,5} (
                \draw[thick, fill = black] (\x + .5, \y) circle (0.2);
            )
        )
        \foreach \x in {2,3,4,5,6,7} (
            \foreach \y in {1,2,3,4,5,6} (
                \draw[thick, fill = white] (\x , \y - .5) circle (0.2);
            )
        )
        \begin{scope}[xshift = -28.5]
    \draw[thick, fill = zcheckcolor] (5,2.5) circle (0.2);
    \draw[thick, fill = zcheckcolor] (4,2.5) circle (0.2);
    \draw[thick, fill = zcheckcolor] (4,3.5) circle (0.2);
    \draw[thick, fill = zcheckcolor] (4.5,2) circle (0.2);
    \draw[thick, fill = zcheckcolor] (3.5,3) circle (0.2);
    \draw[thick, fill = zcheckcolor] (4.5,3) circle (0.2);
    \node[color = black,scale = .5] at (5,2.5) {$1$};
    \node[color = black,scale = .5] at (4,2.5) {$y$};
    \node[color = black,scale = .5] at (4,3.5) {$xy$};
    \node[scale = .5,color = white] at (4.5,2) {$1$};
    \node[scale = .5,color = white] at (3.5,3) {$xy$};
    \node[scale = .5,color = white] at (4.5,3) {$x$};
    \end{scope}
\begin{scope}[xshift = 28.5,yshift = 28.5]
 \draw[thick, fill = xcheckcolor] (5,2.5) circle (0.2);
    \draw[thick, fill = xcheckcolor] (4,2.5) circle (0.2);
    \draw[thick, fill = xcheckcolor] (4,3.5) circle (0.2);
    \draw[thick, fill = xcheckcolor] (4.5,2) circle (0.2);
    \draw[thick, fill = xcheckcolor] (3.5,3) circle (0.2);
    \draw[thick, fill = xcheckcolor] (4.5,3) circle (0.2);
    \draw[fill = xcheckcolor] (3.8,2.8) rectangle ++(.4,.4);
    \node[color = black,scale = .5] at (5,2.5) {$xy$};
    \node[color = black,scale = .5] at (4,2.5) {$x$};
    \node[color = black,scale = .5] at (4,3.5) {$1$};
    \node[scale = .5,color = white] at (4.5,2) {$xy$};
    \node[scale = .5,color = white] at (3.5,3) {$1$};
    \node[scale = .5,color = white] at (4.5,3) {$y$};
    \end{scope}
    \draw[fill = white,color = white] (0,-.5) -- (0,6) -- (1,6) -- (1,-.5) -- cycle;
    \draw[fill = white, color = white] (0,6) -- (7.5,6) -- (7.5,7) -- (0,7) -- cycle;
    \draw[fill = white, color = white] (0,-1) -- (8,-1) -- (8,-.3) -- (0,-.3) -- cycle;
    \draw[fill = white,color = white] (7.3,-.5) -- (7.3,6) -- (8,6) -- (8,-.5) -- cycle;
 \end{tikzpicture} 
         \caption{}\label{subfig:colorcodestabs}
     \end{subfigure}
     \hfill
     \begin{subfigure}[t]{0.25\textwidth}
         \centering
\begin{tikzpicture}[scale = .8,rotate = 270]

        \foreach \x in {2,3,4,5,6,7} (
            \draw[very thick, color = black!60!white] (\x,0) -- (\x, 6);
        )
        \foreach \x in {0,1,2,3,4,5} (
            \draw[very thick, color = black!60!white] (1,\x) -- (7,\x);
        )

\foreach \i in {0,1,2,3,4,5} (
\draw[fill=green,opacity = .2] (2 + \i ,1 - .5 + \i) -- (1.5 + \i,\i) -- (.5 + \i,1 + \i) -- (1+\i,1.5 + \i) -- cycle;
)

\foreach \i in {0,1,2,3,4} (
\draw[fill=green,opacity = .2] (4 + \i ,- .5 + \i) -- (3.5 + \i,\i-1) -- (2.5 + \i, \i) -- (3+\i,.5 + \i) -- cycle;
)

\foreach \i in {1,2} (
\draw[fill=green,opacity = .2] (6 + \i ,- 1.5 + \i) -- (5.5 + \i,\i-2) -- (4.5 + \i,-1+ \i) -- (5+\i,-.5 + \i) -- cycle;
)

\foreach \i in {2,3,4} (
\draw[fill=green,opacity = .2] ( \i , 1.5 + \i) -- (-.5 + \i,\i+1) -- (-1.5 + \i,2 + \i) -- (-1+\i,2.5 + \i) -- cycle;
)


\begin{scope}[xshift = -28.3]
  \foreach \i in {1,2,3,4,5} (
\draw[fill=red,opacity = .2] (2 + \i ,1 - .5 + \i) -- (1.5 + \i,\i) -- (.5 + \i,1 + \i) -- (1+\i,1.5 + \i) -- cycle;
)

\foreach \i in {0,1,2,3,4,5} (
\draw[fill=red,opacity = .2] (4 + \i ,- .5 + \i) -- (3.5 + \i,\i-1) -- (2.5 + \i, \i) -- (3+\i,.5 + \i) -- cycle;
)

\foreach \i in {1,2,3} (
\draw[fill=red,opacity = .2] (6 + \i ,- 1.5 + \i) -- (5.5 + \i,\i-2) -- (4.5 + \i,-1+ \i) -- (5+\i,-.5 + \i) -- cycle;
)

\foreach \i in {3,4} (
\draw[fill=red,opacity = .2] ( \i , 1.5 + \i) -- (-.5 + \i,\i+1) -- (-1.5 + \i,2 + \i) -- (-1+\i,2.5 + \i) -- cycle;
)  
\end{scope}

\begin{scope}[xshift = -56.5]
  \foreach \i in {2,3,4,5} (
\draw[fill=blue,opacity = .2] (2 + \i ,1 - .5 + \i) -- (1.5 + \i,\i) -- (.5 + \i,1 + \i) -- (1+\i,1.5 + \i) -- cycle;
)

\foreach \i in {0,1,2,3,4,5,6} (
\draw[fill=blue,opacity = .2] (4 + \i ,- .5 + \i) -- (3.5 + \i,\i-1) -- (2.5 + \i, \i) -- (3+\i,.5 + \i) -- cycle;
)

\foreach \i in {1,2,3,4} (
\draw[fill=blue,opacity = .2] (6 + \i ,- 1.5 + \i) -- (5.5 + \i,\i-2) -- (4.5 + \i,-1+ \i) -- (5+\i,-.5 + \i) -- cycle;
)

\foreach \i in {4} (
\draw[fill=blue,opacity = .2] ( \i , 1.5 + \i) -- (-.5 + \i,\i+1) -- (-1.5 + \i,2 + \i) -- (-1+\i,2.5 + \i) -- cycle;
)  
\draw[fill=blue, opacity=0.2] (10,-.5) -- (9.5,-1) -- (8.5,0) -- (9,.5) -- cycle;
\end{scope}
        \foreach \x in {1,2,3,4,5,6 } (
            \foreach \y in {0,1,2,3,4,5} (
                \draw[thick, fill = black] (\x + .5, \y) circle (0.2);
            )
        )
        \foreach \x in {2,3,4,5,6,7} (
            \foreach \y in {1,2,3,4,5,6} (
                \draw[thick, fill = white] (\x , \y - .5) circle (0.2);
            )
        )
    
    \draw[fill = white,color = white] (0.5,-.5) -- (0.5,6) -- (1.2,6) -- (1.2,-.5) -- cycle;
    \draw[fill = white, color = white] (0,5.9) -- (7.5,5.9) -- (7.5,7) -- (0,7) -- cycle;
    \draw[fill = white, color = white] (0,-1) -- (8,-1) -- (8,-.3) -- (0,-.3) -- cycle;
    \draw[fill = white,color = white] (7.3,-.5) -- (7.3,6) -- (8,6) -- (8,-.5) -- cycle;
 \end{tikzpicture} 
         \caption{}
         \label{fig:bivascolorcode}
     \end{subfigure}
     \hfill
    \begin{subfigure}[t]{0.35\textwidth}
    \centering
  \begin{tikzpicture}[scale = .8,rotate = 270]

        \foreach \x in {2,3,4,5,6,7} (
            \draw[very thick, color = black!60!white] (\x,0) -- (\x, 6);
        )
        \foreach \x in {0,1,2,3,4,5} (
            \draw[very thick, color = black!60!white] (1,\x) -- (7,\x);
        )

\foreach \i in {0,1,2,3,4,5} (
\draw[fill=green,opacity = .2] (2 + \i ,1 - .5 + \i) -- (1.5 + \i,\i) -- (.5 + \i,1 + \i) -- (1+\i,1.5 + \i) -- cycle;
)

\foreach \i in {0,1,2,3,4} (
\draw[fill=green,opacity = .2] (4 + \i ,- .5 + \i) -- (3.5 + \i,\i-1) -- (2.5 + \i, \i) -- (3+\i,.5 + \i) -- cycle;
)

\foreach \i in {1,2} (
\draw[fill=green,opacity = .2] (6 + \i ,- 1.5 + \i) -- (5.5 + \i,\i-2) -- (4.5 + \i,-1+ \i) -- (5+\i,-.5 + \i) -- cycle;
)

\foreach \i in {2,3,4} (
\draw[fill=green,opacity = .2] ( \i , 1.5 + \i) -- (-.5 + \i,\i+1) -- (-1.5 + \i,2 + \i) -- (-1+\i,2.5 + \i) -- cycle;
)


\begin{scope}[xshift = -28.3]
  \foreach \i in {1,2,3,4,5} (
\draw[fill=red,opacity = .2] (2 + \i ,1 - .5 + \i) -- (1.5 + \i,\i) -- (.5 + \i,1 + \i) -- (1+\i,1.5 + \i) -- cycle;
)

\foreach \i in {0,1,2,3,4,5} (
\draw[fill=red,opacity = .2] (4 + \i ,- .5 + \i) -- (3.5 + \i,\i-1) -- (2.5 + \i, \i) -- (3+\i,.5 + \i) -- cycle;
)

\foreach \i in {1,2,3} (
\draw[fill=red,opacity = .2] (6 + \i ,- 1.5 + \i) -- (5.5 + \i,\i-2) -- (4.5 + \i,-1+ \i) -- (5+\i,-.5 + \i) -- cycle;
)

\foreach \i in {3,4} (
\draw[fill=red,opacity = .2] ( \i , 1.5 + \i) -- (-.5 + \i,\i+1) -- (-1.5 + \i,2 + \i) -- (-1+\i,2.5 + \i) -- cycle;
)  
\end{scope}

\begin{scope}[xshift = -56.5]
  \foreach \i in {2,3,4,5} (
\draw[fill=blue,opacity = .2] (2 + \i ,1 - .5 + \i) -- (1.5 + \i,\i) -- (.5 + \i,1 + \i) -- (1+\i,1.5 + \i) -- cycle;
)

\foreach \i in {0,1,2,3,4,5,6} (
\draw[fill=blue,opacity = .2] (4 + \i ,- .5 + \i) -- (3.5 + \i,\i-1) -- (2.5 + \i, \i) -- (3+\i,.5 + \i) -- cycle;
)

\foreach \i in {1,2,3,4} (
\draw[fill=blue,opacity = .2] (6 + \i ,- 1.5 + \i) -- (5.5 + \i,\i-2) -- (4.5 + \i,-1+ \i) -- (5+\i,-.5 + \i) -- cycle;
)

\foreach \i in {4} (
\draw[fill=blue,opacity = .2] ( \i , 1.5 + \i) -- (-.5 + \i,\i+1) -- (-1.5 + \i,2 + \i) -- (-1+\i,2.5 + \i) -- cycle;
)  
\draw[fill=blue, opacity=0.2] (10,-.5) -- (9.5,-1) -- (8.5,0) -- (9,.5) -- cycle;
\end{scope}
        \foreach \x in {1,2,3,4,5,6 } (
            \foreach \y in {0,1,2,3,4,5} (
                \draw[thick, fill = black] (\x + .5, \y) circle (0.2);
            )
        )
        \foreach \x in {2,3,4,5,6,7} (
            \foreach \y in {1,2,3,4,5,6} (
                \draw[thick, fill = white] (\x , \y - .5) circle (0.2);
            )
        )
    
    \draw[fill = white,color = white] (0.5,-.5) -- (0.5,6) -- (1.2,6) -- (1.2,-.5) -- cycle;
    \draw[fill = white, color = white] (0,5.9) -- (7.5,5.9) -- (7.5,7) -- (0,7) -- cycle;
    \draw[fill = white, color = white] (0,-1) -- (8,-1) -- (8,-.3) -- (0,-.3) -- cycle;
    \draw[fill = white,color = white] (7.3,-.5) -- (7.3,6) -- (8,6) -- (8,-.5) -- cycle;

    \fill[white,opacity = .8] (1,3) -- (1.5,3) -- (4,5.5) -- (4,6) --(1,6) --cycle ;
     \fill[color = white,opacity = .8] (4,5.5) -- (4.5,4) -- (5.5,3) -- (7,2.5) -- (7.5,2.5) -- (7.5,6)--(4,6)--cycle;
     \fill[white,opacity = .8] (7.5,2.5) -- (7,2.5) --  (4.5,0) -- (4.5,-.5) -- (7.5,-.5)--cycle;
      \fill[white,opacity = .8] (4.5,0) -- (4,1.5)--(3,2.5) -- (1.5,3)--(0.7,3) -- (.7,-.5) -- (4.5,-.5) -- cycle;

      \draw (1.5,3) -- (4,5.5) -- (4.5,4) -- (5.5,3) -- (7,2.5) -- (4.5,0) -- (4,1.5) -- (3,2.5) -- cycle;

\foreach \x/\y in {1.5/3, 2.5/4, 3.5/5, 4.5/4, 5.5/3, 6.5/2, 5.5/1, 4.5/0, 3.5/2} (
\draw[thick, fill = black] (\x,\y) circle (0.2);
)

\foreach \x/\y in {2/3.5, 3/4.5, 4/5.5, 5/3.5, 7/2.5, 6/1.5, 5/.5, 4/1.5, 3/2.5}(
\draw[thick, fill = white] (\x,\y) circle (0.2);
)
      
 \end{tikzpicture} 
    \caption{}\label{fig:cutted color code}
    \end{subfigure}
        \caption{The $(6.6.6)$-honeycomb color code as a bivariate bicycle code. In (a), we depict the stabilizer generators for the polynomials $A(x,y) = 1 + x + xy$ and $B(x,y) = 1 + y + xy$. By tiling the plane as in (b), we see that the stabilizer generators induce a $(6.6.6)$- honeycomb color code. On the $6 \times 6$-lattice, this induces a $[[72,4,6]]$ code. In (c), we depict one of many options for pruning the code to an open boundary version. We prune by discarding all stabilizers and qubits that lie completely in the shaded area.  In this specific case, this yields a $[[34,2,4]]$ code. We mention that there are more efficient ways to prune. The presented way behaves well in the context of fold-transversal gates as we will see in Section~\ref{sec: foldtransversal}.}\label{fig:colorcode}
\end{figure}
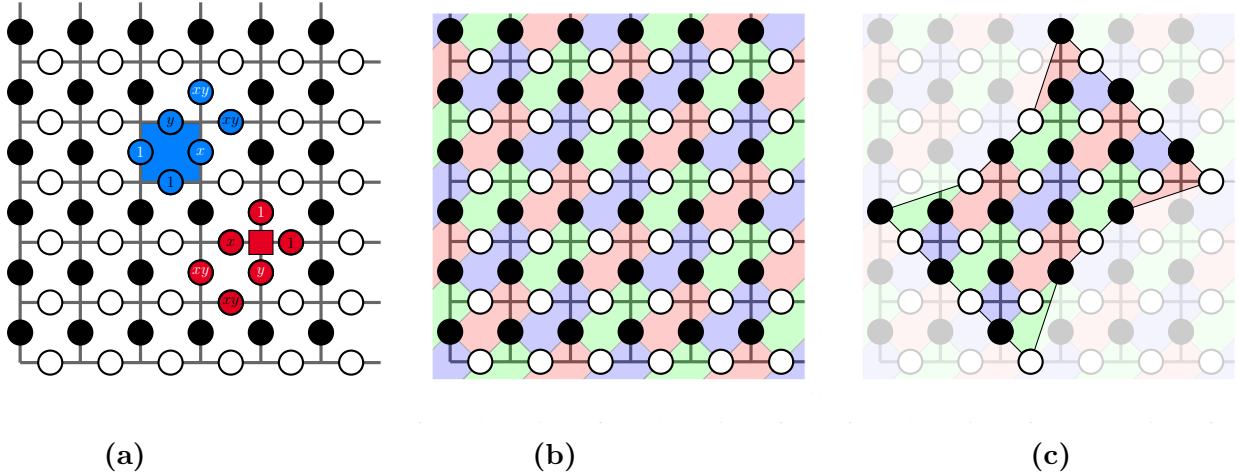

In other words, if $A(x)| (x^{\ell} -1)$ and $B(y) | (y^m -1)$ holds, then we can prune the bivariate bicycle code $\mathcal{Q}$ associated with $A(x)$ and $B(y)$ giving rise to a code that has the same locality properties as $\mathcal{Q}$ on a lattice with open boundary conditions with roughly the same number of qubits, the same distance and half the logical dimension as $\mathcal{Q}$. 

Note that even if $A(x)|(x^{\ell} - 1)$ and $B(y) | (y^m - 1)$ does not hold, one can consider the bivariate bicycle code associated with $\gcd (A(x), x^{\ell} -1)$ and $\gcd (B(y),y^m -1 )$ instead.  Using Lemma~\ref{lem:kergcdlemma}, one can see that this code has the same parameters as the one associated with $A(x)$ and $B(y)$. This new code is at least as local as $\mathcal{Q}$ (but the stabilizers might have a larger support than the ones of $\mathcal{Q}$). Here, all conditions of Proposition~\ref{prop: cutting} are met and we can prune it to a code that does not utilize periodic boundary conditions and has parameters as in Proposition~\ref{prop: cutting}. 

\begin{remark}
We mention that pruning trivial codes as in Lemma~\ref{lem: subcodewithoutperiodicoftensorproduct} leads to non-trivial codes. Indeed, pick integers $\ell$ and $m$ as well as (non-constant) polynomials $A(x)$ and $B(y)$ such that $A(S_\ell)$ and $B(S_m)$ are full rank so that the hypergraph product $\mathcal{C}_{A(S_\ell)}\boxtimes \mathcal{C}_{B(S_m)}^T$ has no logical qubits. Clearly, $A_{red}$ and $B_{red}$ do have kernel, namely of dimension $\deg (A(x))$ resp. $\deg (B(y))$. Consequently, the hypergraph product $\mathcal{C}_{A_{red}} \boxtimes \mathcal{C}_{B_{red}}^T$ has $\deg (A(x)) \cdot\deg (B(y))$ logical qubits. For example, the classical cyclic code on 5 bits associated with $A(x) = B(x) = 1 + x + x^2$ is trivial. Yet, $\mathcal{C}_{A_{red}} \boxtimes \mathcal{C}_{B_{red}}^T$ yields a $[[34,4,3]]$ code. 
\end{remark}

It has been observed that bivariate bicycle codes that are not hypergraph products of cyclic codes, that is, which are induced by polynomials $A(x,y)$ and $B(x,y)$ that are not single variable polynomials can have much more promising parameters~\cite{PhysRevA.88.012311,bravyi2023highthreshold,voss2024trivariatebicyclecodes}. Results as Proposition~\ref{prop: cutting} therefore raise the question if more general bivariate bicycle codes can be pruned. We leave this question for future work but mention that there are examples of bivariate bicycle codes that are not hypergraph product codes but can be pruned.

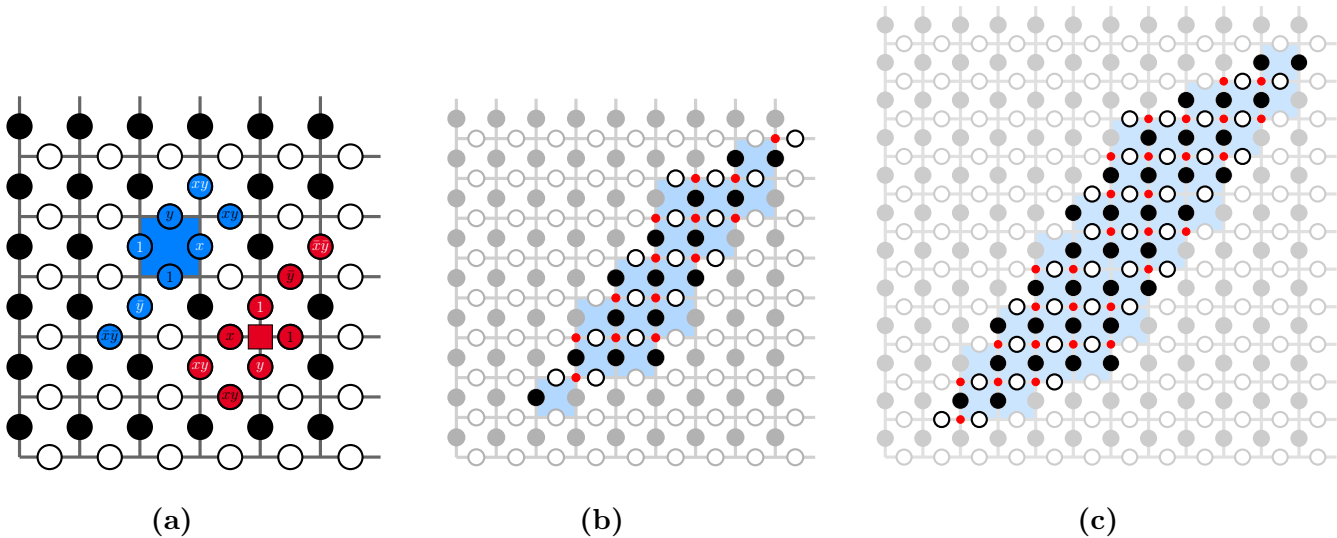
\begin{figure}[h]
     \centering
     \begin{subfigure}[b]{0.25\textwidth}
         \centering

\begin{tikzpicture}[scale = .8,rotate=270]
\draw[color = white] (7.5,0) -- (7.5,1);
          \begin{scope}[xshift = -28]
    \draw[fill = zcheckcolor] (4,2) rectangle ++(1,1);
\end{scope}

        \foreach \x in {2,3,4,5,6,7} (
            \draw[very thick, color = black!60!white] (\x,0) -- (\x, 6);
        )
        \foreach \x in {0,1,2,3,4,5} (
            \draw[very thick, color = black!60!white] (1,\x) -- (7,\x);
        )

        \foreach \x in {1,2,3,4,5,6 } (
            \foreach \y in {0,1,2,3,4,5} (
                \draw[thick, fill = black] (\x + .5, \y) circle (0.2);
            )
        )
        \foreach \x in {2,3,4,5,6,7} (
            \foreach \y in {1,2,3,4,5,6} (
                \draw[thick, fill = white] (\x , \y - .5) circle (0.2);
            )
        )
        \begin{scope}[xshift = -28.5]
    \draw[thick, fill = zcheckcolor] (5,2.5) circle (0.2);
    \draw[thick, fill = zcheckcolor] (4,2.5) circle (0.2);
    \draw[thick, fill = zcheckcolor] (4,3.5) circle (0.2);
    \draw[thick, fill = zcheckcolor] (4.5,2) circle (0.2);
    \draw[thick, fill = zcheckcolor] (3.5,3) circle (0.2);
    \draw[thick, fill = zcheckcolor] (4.5,3) circle (0.2);
    \draw[thick, fill = zcheckcolor] (6,1.5) circle (0.2);
    \draw[thick, fill = zcheckcolor] (5.5,2) circle (0.2);
    \node[color = black,scale = .5] at (5,2.5) {$1$};
    \node[color = black,scale = .5] at (4,2.5) {$y$};
    \node[color = black,scale = .5] at (4,3.5) {$xy$};
    \node[color = black,scale = .5] at (6,1.5) {$\bar{x}\bar{y}$};
    \node[scale = .5,color = white] at (4.5,2) {$1$};
    \node[scale = .5,color = white] at (3.5,3) {$xy$};
    \node[scale = .5,color = white] at (4.5,3) {$x$};
    \node[scale = .5,color = white] at (5.5,2) {$\bar{y}$};
    \end{scope}
\begin{scope}[xshift = 28.5,yshift = 28.5]
 \draw[thick, fill = xcheckcolor] (5,2.5) circle (0.2);
    \draw[thick, fill = xcheckcolor] (4,2.5) circle (0.2);
    \draw[thick, fill = xcheckcolor] (4,3.5) circle (0.2);
    \draw[thick, fill = xcheckcolor] (4.5,2) circle (0.2);
    \draw[thick, fill = xcheckcolor] (3.5,3) circle (0.2);
    \draw[thick, fill = xcheckcolor] (4.5,3) circle (0.2);
    \draw[thick, fill = xcheckcolor] (2.5,4) circle (0.2);
    \draw[thick, fill = xcheckcolor] (3,3.5) circle (0.2);
    \draw[fill = xcheckcolor] (3.8,2.8) rectangle ++(.4,.4);
    \node[color = black,scale = .5] at (5,2.5) {$xy$};
    \node[color = black,scale = .5] at (4,2.5) {$x$};
    \node[color = black,scale = .5] at (4,3.5) {$1$};
    \node[color = black,scale = .5] at (3,3.5) {$\bar{y}$};
    \node[scale = .5,color = white] at (4.5,2) {$xy$};
    \node[scale = .5,color = white] at (3.5,3) {$1$};
    \node[scale = .5,color = white] at (4.5,3) {$y$};
    \node[scale = .5,color = white] at (2.5,4) {$\bar{x}\bar{y}$};
    \end{scope}

 \end{tikzpicture}

 \caption{}\label{subfig:fullcodestabs}
\end{subfigure}
 \hfill
 \begin{subfigure}[b]{0.25\textwidth}
         \centering
\begin{tikzpicture}[scale = .53,rotate = 270]

\draw[color = white] (10.75,0) -- (10.75,1);

\pgfmathsetmacro{\n}{10}
\pgfmathsetmacro{\m}{\n - 2}
\pgfmathsetmacro{\o}{\n - 1}
\pgfmathsetmacro{\p}{\n - 4}
\pgfmathsetmacro{\opa}{1}

\foreach \i\j in {3/7,4/5,4/6,4/7,5/5,5/6,6/4,6/5,7/3,7/4,7/5,8/3,8/4,9/2}
{
    \draw[color = black, fill = zcheckcolor] (\i,\j) rectangle ++(-1,1);
}

                \foreach \x in {2,...,\n} (
            \draw[very thick, color = black!60!white, opacity = \opa] (\x,0) -- (\x, \o);
        )
        \foreach \x in {0,...,\m} (
            \draw[very thick, color = black!60!white, opacity = \opa] (1,\x) -- (\n,\x);
        )

        \foreach \x in {1,...,\o} (
            \foreach \y in {0,...,\m} (
                \draw[thick, fill = black, opacity = \opa] (\x + .5, \y) circle (0.2);
            )
        )
        \foreach \x in {2,...,\n} (
            \foreach \y in {1,...,\o} (
                \draw[thick, fill = white,opacity = \opa] (\x , \y - .5) circle (0.2);
            )
        )

\fill[white,opacity = .7] (0.5,-.5) -- (\n + .5,-.5) -- (\n + .5,\o + .5) -- (0.5,\o + .5) --cycle;

\foreach \i\j in {2.5/8, 2.5/7, 3.5/6, 3.5/7, 4.5/5, 4.5/6, 5.5/4, 5.5/5, 5.5/6,
6.5/4, 6.5/5, 7.5/4, 7.5/5, 7.5/3, 8.5/2}
{
    \draw[fill = black] (\i,\j) circle (.2);
}
\foreach \i\j in {2/8.5, 
3/7.5,3/6.5, 3/5.5, 
4/6.5,4/5.5,
5/6.5,5/5.5,5/4.5,
6/5.5,6/4.5,
7/4.5,7/3.5,
8/2.5,8/3.5}
{
    \draw[thick, fill = white] (\i,\j) circle (.2);
}

\foreach \i\j in {2/8,3/7,3/6,4/7,4/6,4/5,5/5,5/6,6/4,6/5,7/3,7/4,7/5,8/3}
{
    \draw[color = red, fill =red] (\i,\j) circle (.1);
}

        \end{tikzpicture}
         \caption{}
         \label{fig:smallprune}
     \end{subfigure}
     \hfill
    \begin{subfigure}[b]{0.35\textwidth}
    \centering
  \begin{tikzpicture}[scale = .5,rotate = 270]

  \draw[color = white] (13.8,0) -- (13.8,1);

\pgfmathsetmacro{\n}{13}
\pgfmathsetmacro{\m}{\n - 2}
\pgfmathsetmacro{\o}{\n - 1}
\pgfmathsetmacro{\p}{\n +1}
\pgfmathsetmacro{\opa}{1}

\foreach \i\j in {
3/9,
4/7,4/8,4/9,
5/5,5/6,5/7,5/8,
6/5,6/6,6/7,
7/4,7/5,7/6,7/7,
8/3,8/4,8/5,8/6,
9/3,9/4,9/5,
10/2,10/3,10/4,10/5,
11/1,11/2,11/3,11/4,
12/1,12/2
}
{
    \draw[color = black, fill = zcheckcolor] (\i,\j+1) rectangle ++(-1,1);
}

                \foreach \x in {2,...,\n} (
            \draw[very thick, color = black!60!white, opacity = \opa] (\x,0) -- (\x, \o);
        )
        \foreach \x in {0,...,\m} (
            \draw[very thick, color = black!60!white, opacity = \opa] (1,\x) -- (\n,\x);
        )

        \foreach \x in {1,...,\o} (
            \foreach \y in {0,...,\m} (
                \draw[thick, fill = black, opacity = \opa] (\x + .5, \y) circle (0.2);
            )
        )
        \foreach \x in {2,...,\n} (
            \foreach \y in {1,...,\o} (
                \draw[thick, fill = white,opacity = \opa] (\x , \y - .5) circle (0.2);
            )
        )

\fill[white,opacity = .8] (0.5,-.5) -- (\n + .5,-.5) -- (\n + .5,\o + .5) -- (0.5,\o + .5) --cycle;

\foreach \i\j in {
2.5/9,2.5/10,
3.5/7,3.5/8,3.5/9,
4.5/6,4.5/7,4.5/8,
5.5/5,5.5/6,5.5/7,5.5/8,
6.5/4,6.5/5,6.5/6,6.5/7,
7.5/4,7.5/5,7.5/6,
8.5/3,8.5/4,8.5/5,8.5/6,
9.5/2,9.5/3,9.5/4,9.5/5,
10.5/2,10.5/3,10.5/4,10.5/5,
11.5/1,11.5/2
}
{
    \draw[fill = black] (\i,\j+1) circle (.2);
}
\foreach \i\j in {
3/8.5,3/9.5,
4/5.5,4/6.5,4/7.5,4/8.5,
5/5.5,5/6.5,5/7.5,5/8.5,
6/4.5,6/5.5,6/6.5,6/7.5,
7/4.5,7/5.5,7/6.5,
8/3.5,8/4.5,8/5.5,8/6.5,
9/2.5,9/3.5,9/4.5,9/5.5,
10/2.5,10/3.5,10/4.5,
11/1.5,11/2.5,11/3.5,
12/.5,12/1.5
}
{
    \draw[thick, fill = white] (\i,\j+1) circle (.2);
}

\foreach \i\j in {
3/9,3/8,
4/6,4/7,4/8,4/9,
5/5,5/6,5/7,5/8,
6/5,6/6,6/7
7/4,7/5,7/6,7/7,
8/3,8/4/8/5,8/6,
9/3,9/4,9/5,
10/2,10/3,10/4,10/5,
11/1,11/2,11/3,
12/1
}
{
    \draw[color = red, fill =red] (\i,\j+1) circle (.1);
}

        \end{tikzpicture}
    \caption{}\label{fig:bigprune}
    \end{subfigure}
        \caption{Pruning the code associated with the polynomials $A(x,y) = 1 + x + y^{-1} + xy$ and $B(x,y) = 1 + y +xy + x^{-1}y^{-1}$. In~\ref{subfig:fullcodestabs}, we depict examples of $X$- and $Z$-type stabilizers where we write $\bar{x} = x^{-1}$ and $\bar{y} = y^{-1}$ for better visibility. In ~\ref{fig:smallprune} and~\ref{fig:bigprune}, we depict pruned versions of it resulting in codes with parameters $[[30,2,4]]$ and $[[66,2,6]]$, respectively. }\label{fig:colorcode}
\end{figure}

\begin{example}\label{ex:colorcode}
    Consider $A(x,y) = 1 + x + xy$ and $B(x,y) = 1 + y + xy$. It turns out that for $\ell,m$ multiples of 3, this defines the $(6.6.6)$-honeycomb color code~\cite{Bombin_2006}. In Figure~\ref{subfig:colorcodestabs} and~\ref{fig:bivascolorcode} we show the stabilizer generators and how one can see that they are the stabilizer generators for the $(6.6.6)$-honeycomb color code.  
There are many known ways to prune this code yielding codes on a lattice with open boundary conditions with interesting parameters, we depict one example in Figure~\ref{fig:cutted color code}.
\end{example}

\begin{example}\label{ex: fullcode}
The pruning of the color code in Example~\ref{ex:colorcode} relies on the intuition that the code is defined by the underlying hexagonal lattice.
To demonstrate that pruning is also possible for less intuitive examples, we show in Figure~\ref{fig:bigprune} that also the code associated with $A(x,y) = 1 + x + xy + y^{-1}$ and $B(x,y) = 1 + y + xy + (xy)^{-1}$ can be pruned.
We note that these generalize to a code family with fixed number of 2 logical qubits and distance scaling with the patch size. The shown examples were constructed using a computer search over certain convex regions in the lattice.
\end{example}

\section{Fold-transversal gates}\label{sec: foldtransversal}
We now want to turn our attention to the relation of recent proposals of fault-tolerant gates for bivariate bicycle codes and the ways of pruning them investigated in Section~\ref{sec: towardsopenboundary}. One proposal are so-called \textit{fold-transversal gates}~\cite{Moussa_2016, breuckmannFoldTransversalCliffordGates2024} which have been investigated in the context of hypergraph product codes~\cite{Quintavalle_2023} and in the context of bivariate bicycle codes~\cite{eberhardt2024logicaloperatorsfoldtransversalgates}. In this section, we briefly review the fold-transversal gates on an intuitive level and explain how they also induce fold-transversal gates for their pruned versions that we saw in Section~\ref{sec: towardsopenboundary}.

\begin{definition}
    Let $\mathcal{Q}$ be a CSS code on a set of physical qubits $\lbrace 1, \dots, n \rbrace$ induced by $X$-check and $Z$-check matrices $H_X$ and $H_Z$. An \textit{automorphism} of $\mathcal{Q}$ is given by a $n \times n$ permutation matrix $\sigma_n$ such that there exists permutation matrices $\sigma_X$ and $\sigma_Z$ with $H_X = \sigma_{X} H_X \sigma_n $ and $ H_Z = \sigma_{Z} H_Z \sigma_n.$ When $H_X$ and $H_Z$ have equally many rows, we say that $\sigma_n$ is a \textit{$ZX$-duality} if there are permutation matrices $\tau_{X}$ and $\tau_{Z}$ such that $H_Z = \tau_{X} H_X \sigma_n $ and $H_X = \tau_{Z} H_Z \sigma_n.$
\end{definition}

In other words, an automorphism is a permutation of the qubits that induces a one-to-one permutation of the $X$-type stabilizers and $Z$-type stabilizers. A $ZX$-duality is a permutation of the qubits that maps $X$-type stabilizers to $Z$-type stabilizers and vice versa. 

Automorphisms and $ZX$ dualities can be used to construct so-called fold-transversal gates~\cite{breuckmannFoldTransversalCliffordGates2024}. Given a $ZX$ duality $\tau$, a fold transversal gate with respect to $\tau$ is any unitary that can be realized as a product of single qubit gates and two-qubit gates between qubit pairs $i, \tau (i)$. For example, any $ZX$-duality $\tau$ induces a \textit{Hadamard-type logical gate} that applies  
\begin{equation*}
   \bigotimes_{i = 1, \dots, n, i < \tau_n(i)} \op{SWAP}_{i ,\tau(i)} \bigotimes_{i = 1, \dots, n} H_i,
\end{equation*}
see~\cite{breuckmannFoldTransversalCliffordGates2024} for details. 

For bivariate bicycle codes that are symmetric in the sense that $\ell = m$ and $A(x,y) = B(y,x)$, there is a $ZX$ duality that takes the two sub-lattices of horizontal and vertical qubits and reflects the qubits in each of them along the anti-diagonal. This has been investigated for hypergraph products of two copies of the same classical (and not  necessarily cyclic) code in~\cite{Quintavalle_2023} under the name \textit{digital twin partition} and for bivariate bicycle codes in~\cite{eberhardt2024logicaloperatorsfoldtransversalgates}. For convenience, we visualize this in Figure~\ref{subfig:zxduality}. We will write $\pi$ for the permutation of the qubits associated with it.

\newcommand{\drawlattice}{
        \foreach \x in {1,2,3,4,5,6} (
            \draw[thick, color = black!60!white] (\x,-1) -- (\x, 5);
        )
        \foreach \x in {-1,0,1,2,3,4} (
            \draw[thick, color = black!60!white] (1,\x) -- (7,\x);
        )
            \foreach \x in {1,2,3,4,5,6} (
            \foreach \y in {0,1,2,3,4,5} (
                \draw[fill = black!20!white] (\x , \y - .5) circle (0.2);
            )
                   \foreach \x in {1,2,3,4,5,6} (
            \foreach \y in {-1,0,1,2,3,4} (
                \draw[fill = white] (\x + .5, \y) circle (0.2);
            )
        )
}

\begin{figure}
\begin{subfigure}[t]{0.3\textwidth}
\centering

\begin{tikzpicture}[scale = .7]
        \foreach \x in {1,2,3,4,5,6} (
            \draw[very thick, color = black!60!white] (\x,-1) -- (\x, 5);
        )
        \foreach \x in {-1,0,1,2,3,4} (
            \draw[very thick, color = black!60!white] (1,\x) -- (7,\x);
        )
\draw[very thick, color = green!60!black] (7,-1.5) -- (.5,5);
            \foreach \x in {1,2,3,4,5,6} (
            \foreach \y in {0,1,2,3,4,5} (
                \draw[thick, fill = black!20!white] (\x , \y - .5) circle (0.2);
            )
        )
        \draw[very thick, color = green!60!black,<->] (3,3.5) .. controls (2.2,3.3) .. (2,2.5);
                   \foreach \x in {1,2,3,4,5,6} (
            \foreach \y in {-1,0,1,2,3,4} (
                \draw[thick, fill = white] (\x + .5, \y) circle (0.2);
            )
        )
\draw[very thick, color = green!60!black, <->] (3.5,0) .. controls (4.9,.6) .. (5.5,2);
\end{tikzpicture}

\caption{}\label{subfig:zxduality}
\end{subfigure}
\hfill
\begin{subfigure}[t]{0.3\textwidth}
\centering

\begin{tikzpicture}[scale = .7]
        \foreach \x in {1,2,3,4,5,6} (
            \draw[very thick, color = black!60!white] (\x,-1) -- (\x, 5);
        )
        \foreach \x in {-1,0,1,2,3,4} (
            \draw[very thick, color = black!60!white] (1,\x) -- (7,\x);
        )
\draw[very thick, color = green!60!black] (7,-1.5) -- (.5,5);
            \foreach \x in {1,2,3,4,5,6} (
            \foreach \y in {0,1,2,3,4,5} (
                \draw[thick, fill = black!20!white] (\x , \y - .5) circle (0.2);
            )
        )
        \draw[very thick, color = green!60!black] (3,3.5) .. controls (2.2,3.3) .. (2,2.5);
                   \foreach \x in {1,2,3,4,5,6} (
            \foreach \y in {-1,0,1,2,3,4} (
                \draw[thick, fill = white] (\x + .5, \y) circle (0.2);
            )
        )
\draw[very thick, color = green!60!black] (3.5,0) .. controls (4.9,.6) .. (5.5,2);

        \foreach \x in {-1,0,1,2,3,4}(
\node[scale = .5, color = green!60!black] at (5 - \x,.5 + \x) {$S^\dagger$};
)
\foreach \x in {-2,-1,0,1,2,3}(
\node[scale = .5, color = green!60!black] at (4.5 - \x, 1+ \x) {$S$};
)

\node[scale = .6, color = green!60!black]  at (2.5,3.6) {$CZ$};

\node[scale = .6, color = green!60!black]  at (5.52,1.4) {$CZ$};
\end{tikzpicture}

\caption{}
\label{subfig:phasetypegate}
\end{subfigure}
\hfill
\begin{subfigure}[t]{0.3\textwidth}
\centering

\begin{tikzpicture}[scale = .7]
    \fill[green!80!black,opacity = .2] (0.75,-1.5) -- (.75,4.8) -- (7,4.8) -- (7,3.25) -- (2.25,3.25) -- (2.25,-1.5)  -- cycle;
    \fill[white] (1.25,-1.25)--  (1.25,3.25) --(1.75,3.25)  -- (1.75,-1.25) -- cycle;

\fill[white] (2.25,3.75)--  (2.25,4.25) --(6.75,4.25)  -- (6.75,3.75) -- cycle;

    \fill[white] (1.25,3.75) -- (1.25,4.25) -- (1.75,4.25) -- (1.75,3.75) -- cycle;
        \foreach \x in {1,2,3,4,5,6} (
            \draw[very thick, color = black!60!white] (\x,-1) -- (\x, 5);
        )
        \foreach \x in {-1,0,1,2,3,4} (
            \draw[very thick, color = black!60!white] (1,\x) -- (7,\x);
        )
            \foreach \x in {1,2,3,4,5,6} (
            \foreach \y in {0,1,2,3,4,5} (
                \draw[thick, fill = black!20!white] (\x , \y - .5) circle (0.2);
            )
        )
                   \foreach \x in {1,2,3,4,5,6} (
            \foreach \y in {-1,0,1,2,3,4} (
                \draw[thick, fill = white] (\x + .5, \y) circle (0.2);
            )
        )
        \draw[thick,densely dotted] (5.5,-1) circle (.3);
\draw[thick,densely dotted] (6.5,-1) circle (.3);
\draw[thick,densely dotted] (5.5,0) circle (.3);
\draw[thick,densely dotted] (6.5,0) circle (.3);
\draw[thick,densely dotted] (2,4.5) circle (.3);
\draw[thick,densely dotted] (1,4.5) circle (.3);
\draw[thick,densely dotted] (2,3.5) circle (.3);
\draw[thick,densely dotted] (1,3.5) circle (.3);
\end{tikzpicture}

\setcounter{subfigure}{3}
\caption{}
\label{subfig:cuttedsublattice}
\end{subfigure}
\hfill
\vspace{1em}
\begin{subfigure}[t]{\textwidth}
\centering

\begin{tikzpicture}[scale = .5]
    \drawlattice
    
    \foreach \i in {1.5,2.5,4.5,5.5}(
    \fill[color = zcheckcolor] (\i,-1) circle (.2);
    )

    \fill[color  = xcheckcolor] (5.3,-1) -- (5.7,-1) arc(0:135:0.2) --cycle;
    \draw[thick,densely dotted] (5.5,-1) circle (.3);
    \foreach \j in {1,2,4}(
    \fill[color = xcheckcolor] (5.5,\j) circle (.2);
    )
\begin{scope}[xshift = 18em]
    \drawlattice

\foreach \i in {1.5,3.5,4.5,6.5}(
\fill[color = zcheckcolor] (\i,-1) circle (.2);
)

\fill[color  = xcheckcolor] (6.3,-1) -- (6.7,-1) arc(0:135:0.2) --cycle;
\draw[thick,densely dotted] (6.5,-1) circle (.3);
\foreach \j in {1,2,4}(
\fill[color = xcheckcolor] (6.5,\j) circle (.2);
)

\end{scope}

\begin{scope}[xshift = 36em]
    \drawlattice
    \foreach \i in {1.5,2.5,4.5,5.5}(
    \fill[color = zcheckcolor] (\i,0) circle (.2);
    )

    \fill[color  = xcheckcolor] (5.3,0) -- (5.7,0) arc(0:135:0.2) --cycle;
    \draw[thick,densely dotted] (5.5,0) circle (.3);
    \foreach \j in {1,3,4}(
    \fill[color = xcheckcolor] (5.5,\j) circle (.2);
    )

\end{scope}

\begin{scope}[xshift = 54em]
    \drawlattice
    \foreach \i in {1.5,3.5,4.5,6.5}(
    \fill[color = zcheckcolor] (\i,0) circle (.2);
    )

    \fill[color  = xcheckcolor] (6.3,0) -- (6.7,0) arc(0:135:0.2) --cycle;
    \draw[thick,densely dotted] (6.5,0) circle (.3);
    \foreach \j in {1,3,4}(
    \fill[color = xcheckcolor] (6.5,\j) circle (.2);
    )
\end{scope}

\begin{scope}[yshift = -18em]
\drawlattice
\foreach \i in {2,3,5,6}(
\fill[color = xcheckcolor] (\i,4.5) circle (.2);
)

\fill[color  = zcheckcolor] (1.8,4.5) -- (2.2,4.5) arc(0:-135:0.2) --cycle;
\draw[thick,densely dotted] (2,4.5) circle (.3);
\foreach \j in {-.5,1.5,2.5}(
\fill[color = zcheckcolor] (2,\j) circle (.2);
)
\end{scope}

\begin{scope}[xshift = 18em,yshift = -18em]
\drawlattice
\foreach \i in {1,3,4,6}(
\fill[color = xcheckcolor] (\i,4.5) circle (.2);
)

\fill[color  = zcheckcolor] (0.8,4.5) -- (1.2,4.5) arc(0:-135:0.2) --cycle;
\draw[thick,densely dotted] (1,4.5) circle (.3);
\foreach \j in {-.5,1.5,2.5}(
\fill[color = zcheckcolor] (1,\j) circle (.2);
)
\end{scope}

\begin{scope}[xshift = 36em,yshift = -18em]
\drawlattice
\foreach \i in {2,3,5,6}(
\fill[color = xcheckcolor] (\i,3.5) circle (.2);
)

\fill[color  = zcheckcolor] (1.8,3.5) -- (2.2,3.5) arc(0:-135:0.2) --cycle;
\draw[thick,densely dotted] (2,3.5) circle (.3);
\foreach \j in {-.5,0.5,2.5}(
\fill[color = zcheckcolor] (2,\j) circle (.2);
)
\end{scope}

\begin{scope}[xshift = 54em,yshift = -18em]
\drawlattice
\foreach \i in {1,3,4,6}(
\fill[color = xcheckcolor] (\i,3.5) circle (.2);
)

\fill[color  = zcheckcolor] (0.8,3.5) -- (1.2,3.5) arc(0:-135:0.2) --cycle;
\draw[thick,densely dotted] (1,3.5) circle (.3);
\foreach \j in {-.5,0.5,2.5}(
\fill[color = zcheckcolor] (1,\j) circle (.2);
)
\end{scope}
    
\end{tikzpicture}

\setcounter{subfigure}{2}
\caption{}\label{subfig:logicalbasisalaquintavalle}
\end{subfigure}
\hfill
\caption{In (a), we depict the $ZX$ duality $\pi$. For any symmetric bivariate bicycle code, it induces a logical operation by applying $S$, $S^\dagger$, and $CZ$ on the physical qubits as depicted in (b). In (c), we depict a basis of logical operators for the bivariate bicycle code associated with $A(x) = 1 + x + x^2$ and $B(y) = 1 + y + y^2$ using the construction from~\cite{Quintavalle_2023}. It has the desirable property that each pair of logical $X$ and $Z$ operators overlaps on no or exactly one physical qubits. One can thus think of the highlighted physical qubits on which logicals intersect as labeling the logical qubits. In (d) we show again the pruned versions as presented in Lemma~\ref{lem: subcodewithoutperiodicoftensorproduct}. Here, it becomes clear how the logical dimension drops to half the initial logical dimension: Exactly the logical qubits whose labeling physical qubit gets discarded are lost. Combining (b), (c) and (d) one can nicely see that by construction, the phase-type gate induces a logical gate on the pruned code as well, namely the same logical gate restricted to the remaining logical qubits.}
\end{figure}
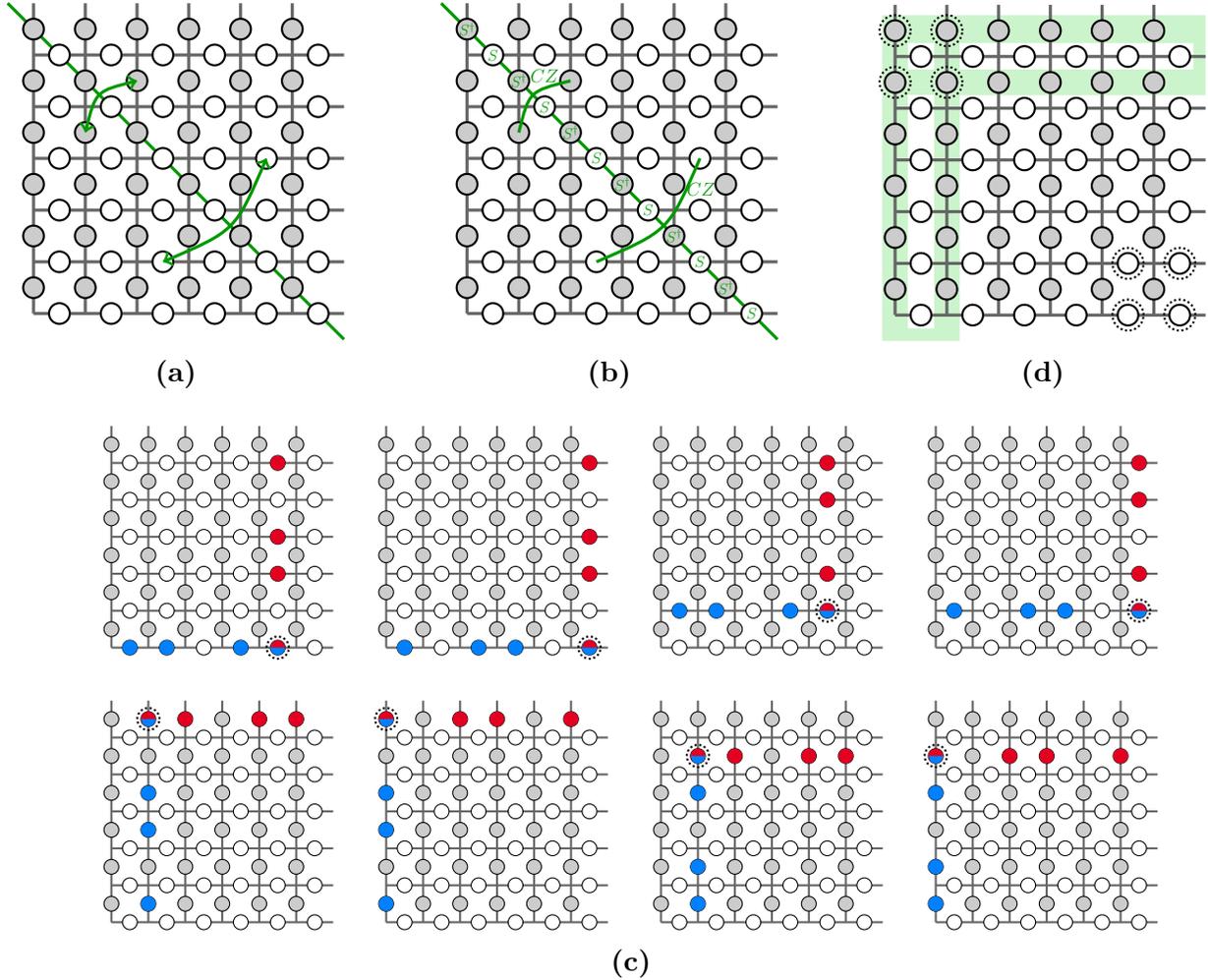

For hypergraph products of two copies of the same classical codes as well as for symmetric bivariate bicycle codes, the $ZX$-duality $\pi$ induces a phase type fold-transversal gate which acts by applying $S$ to all horizontal qubits on the diagonal, $S^\dagger$ to all the vertical qubits on the diagonal and $CZ$ gates to the pairs of qubits that make up the two-qubit orbits of $\pi$, see Figure~\ref{subfig:phasetypegate} for a visualization~\cite{Quintavalle_2023, eberhardt2024logicaloperatorsfoldtransversalgates}. For bivariate bicycle codes induced by single-variable polynomials $A(x)$ and $B(y)$, the action of this gate on a logical level can be nicely understood using the results from~\cite{Quintavalle_2023}. We only mention the key points and refer for technical details to~\cite{Quintavalle_2023}. It turns out that any hypergraph product of a pair of classical linear codes admits a basis of logical $X$- and $Z$-operators such that each pair of $X$- and $Z$- operators overlaps on no or exactly one qubit. We depict the situation for the bivariate bicycle code induced by $A(x) = 1 + x + x^2$ and $B(y) = 1 + y + y^2$ in Figure~\ref{subfig:logicalbasisalaquintavalle} to give some intuition. The choice of such a basis of logical operator fixes the choice of a basis of the logical qubits: In fact, one can think of the logical qubits being labeled by the sub-lattice of intersection points of the $X$- and $Z$-operators. In this way, it has been demonstrated in~\cite{Quintavalle_2023} that the phase type gate is a unitary on the logical level that applies an $S$ ($S^\dagger$) gate to all logical qubits whose label qubit is a horizontal (vertical) qubit on the diagonal and $CZ$ to all pairs of logical qubits whose label qubits are mapped to one another under the permutation $\pi$.

It is interesting to ask what the influence of pruning is. The following observation is immediate looking at Figure~\ref{fig:cuttingsubfigure} and~\ref{fig:cutted color code}.
\begin{proposition}
    The $ZX$ duality $\pi$ is compatible with the pruning as done in Lemma~\ref{lem: subcodewithoutperiodicoftensorproduct} and Example~\ref{ex:colorcode} in the sense that its restriction to the remaining physical qubits still yields a unitary on the logical level.
\end{proposition}

\begin{remark}
Another simple way of constructing an automorphism is by shifting the lattice, that is, identifying the qubits of the two lattices with monomials in $\mathbb{F}_2[x,y]/(x^\ell -1, y^m - 1)$, multiplying by a monomial, see~\cite{eberhardt2024logicaloperatorsfoldtransversalgates} for more details. We mention that automorphisms constructed like this are in general not compatible with the way we pruned codes. In~\cite{Quintavalle_2023}, another type of $ZX$-duality for hypergraph product codes induced by self-dual classical codes has been investigated. While a variation of this yields a fold-transversal gate for all symmetric bivariate bicycle codes induced by single-variable polynomials (compare also Lemma~\ref{lem: transpose of cyclic code}), it is not compatible with pruning as in Lemma~\ref{lem: subcodewithoutperiodicoftensorproduct}. 
\end{remark}

In the way of thinking of the logical qubits of the bivariate bicycle code associated with $A(x)$ and $B(y)$ being labeled by the intersection points of logical operators as in Figure~\ref{subfig:logicalbasisalaquintavalle}, one can nicely understand how the logical dimension of the pruned version is half of the dimension of the initial bivariate bicycle code: In Figure~\ref{subfig:cuttedsublattice} one can observe that exactly the qubits survive whose label qubits not being pruned away, the logical qubits corresponding to label qubits that are being discarded do not. In this way one can also nicely understand the logical action of the phase type gate on the pruned version which is exactly the action of the logical operation on the initial code restricted to the logical qubits that survive. 

We can also see that, by the particular construction, for the pruning as in Example~\ref{ex:colorcode} the restriction of the phase type gate to the remaining qubits also yields a valid logical operation.

\section{Conclusion and outlook}\label{sec: outlook}

In this work, we reviewed the bivariate bicycle codes which are promising candidates for qLDPC codes to be implemented on near-term hardware. While the bivariate bicycle codes have checks that can be forced to be local on a 2D lattice, they by construction require 2D connectivity on a lattice with periodic boundary conditions which often imposes a major roadblock for physical implementation. The main purpose of this work is to raise the question if bivariate bicycle codes can be transformed into codes that have the same locality properties but on a lattice with open boundary conditions and similarly promising code parameters. To set the stage for this question, we explained how this is always possible for bivariate bicycle codes that are induced by single variable polynomials $A(x)$ and $B(y)$, that is, bivariate bicycle codes that are hypergraph products of classical cyclic codes. We also demonstrated that for certain examples of bivariate bicycle codes that are not hypergraph products, this is possible by drawing a connection to color codes. Apart from that, this work opens the door to a variety of questions:
\begin{enumerate}
\item First and foremost, we ask if one can prune more general bivariate bicycle codes, in particular, bivariate bicycle codes with promising code parameters as constructed for example in~\cite{bravyi2023highthreshold}. Similar constructions would yield codes that are much more likely to be implemented on real hardware in the near future. 
\item The procedure of pruning is only one way of transforming bivariate bicycle codes into codes with similar locality properties on a lattice with open boundary conditions. In~\cite{PRXQuantum.5.030328, liang2024operatoralgebraalgorithmicconstruction}, the authors investigate the \textit{boundary gauge operators} that arise by pruning bivariate bicycle codes. Introducing some of these boundary gauge operators additionally as stabilizers then might lead to a code with local checks and favorable parameters.
\item Due to their symmetric structure, many bivariate bicycle codes~\cite{eberhardt2024logicaloperatorsfoldtransversalgates} support fold-transversal gates which are proposals for fault-tolerant implementations of certain Clifford gates. We have seen in Section~\ref{sec: foldtransversal} that some of these fold-transversal gates are compatible with our way of pruning and induce fold-transversal gates on the pruned versions. When finding ways of pruning more general bivariate bicycle codes, it is interesting to ask if any structural insights from the bivariate bicycle codes carry over to the pruned versions. 
\item In~\cite{xu2024fastparallelizablelogicalcomputation}, related ways of pruning hypergraph product codes have been investigated in the context of implementing measurements of Pauli operators on hypergraph product codes. A suitable way of pruning general bivariate bicycle codes would open the door to apply a similar strategy there. 
\end{enumerate}

\bibliographystyle{alpha}
\bibliography{qLDPC}
\end{document}